\documentclass[12pt,preprint]{aastex}

\shorttitle{Steady Line-Driven Disk Winds in CVs}

\shortauthors{Pereyra et~al.}

\begin{document}

\title{On the Steady Nature of Line-Driven Disk Winds:
       Application to Cataclysmic Variables}

\author{Nicolas A. Pereyra\altaffilmark{1}, David A. Turnshek,
        and D. John Hillier}
\affil{University of Pittsburgh, Department of Physics and Astronomy,
       3941 O'Hara ST, Pittsburgh, PA 15260}
\email{pereyra@bruno.phyast.pitt.edu,
       turnshek@quasar.phyast.pitt.edu,
       jdh@galah.phyast.pitt.edu}

\altaffiltext{1}{Also at:
                 Universidad Mar\'\i tima del Caribe,
                 Departamento de Ciencias B\'asicas,
                 Catia la Mar, Venezuela}

\begin{abstract}
We apply the semi-analytical analysis of the steady nature of
line-driven winds presented in two earlier papers to disk winds
driven by the flux distribution of a standard \citet{sha73}
disk for typical cataclysmic variable (CV) parameters.
We find that the wind critical point tends to be closer to the disk
surface towards the inner disk regions.
Our main conclusion,
however,
is that a line-driven wind,
arising from a steady disk flux distribution of a standard
Shakura-Sunyaev disk capable of locally supplying the corresponding
mass flow,
is steady.
These results confirm the findings of an earlier paper that studied
``simple''
flux distributions that are more readily analyzable than those
presented here.
These results are consistent with the steady velocity nature of outflows
observationally inferred for both CVs and quasi-stellar objects
(QSOs).
We find good agreement with the 2.5D CV disk wind models of Pereyra and
collaborators.
These results suggest that the likely scenario to account for the wind
outflows commonly observed in CVs is the line-driven accretion disk wind
scenario,
as suggested early-on by \citet{cor82}.
For QSOs,
these results show that the line-driven accretion disk wind continues to
be a promising scenario to account for the outflows detected in 
broad absorption line (BAL) QSOs,
as suggested early-on by \citet{tur84},
and analyzed in detail by \citet{mur95}.

\end{abstract}


\section{Introduction}
\label{sec_introduction}

As discussed in
\citet[][hereafter Paper~I]{per04}
and
\citet[][hereafter Paper~II]{per05},
accretion disks are commonly believed to be present in both cataclysmic
variables
(CVs)
and quasi-stellar objects and active galactic nuclei
(QSOs/AGNs).
A property that CVs and QSOs have in common is that both types of
objects sometimes present blue-shifted absorption troughs in UV
resonance lines,
giving direct observational evidence for an outflowing wind.
Both types of objects also show the existence of a persistent velocity
structure in their absorption troughs
(when present)
over significantly long time scales (Paper~I).

In order for a disk wind to account for the wide/broad resonance line
absorption structures observed in many CVs and QSOs,
it must be able to account for the steady velocity structure that is
observed.
The 2.5D time-dependent disk wind models of Pereyra and collaborators,
both for CVs and QSOs,
have steady disk wind solutions
\citep{per97a,per97b,per00,hil02,per03}.
The earlier 1D disk wind models of Murray and collaborators also find
steady disk wind solutions
\citep{mur95,mur96,chi96,mur98}.

However,
in the literature it has been argued that
line-driven disk winds are ``intrinsically unsteady''
because of the increasing gravity along the streamlines at wind base
that is characteristic of disk winds
(see Paper~I; Paper~II).
Since the steady nature of CV and QSO wind flows is an observational
constraint,
whether line-driven disk winds are steady is a significant issue.
Our objective is to study this issue through a semi-analytical method
independent of our previous 2.5D numerically intensive simulation
efforts
\citep[e.g.,][]{per97a,per00}.

In Paper~I we showed that an increase in gravity at wind base,
that is characteristic of disk winds,
does {\it not} imply an unsteady wind solution.
We also developed mathematically ``simple'' models that mimic the disk
environment and we showed that
{\it line-driven disk winds can be steady}.
In Paper~II we extended the concepts presented in Paper~I,
and discussed in detail many aspects that were mentioned in Paper~I.

The goal of this paper is to apply the concepts introduced in Paper~I
and Paper~II to more realistic models that implement the exact flux
distribution above a standard accretion disk.
We compare and find good agreement with the steady CV disk wind
solutions found through numerically intensive calculations by
\citet{per97a}
and
\citet{per00}.

In \S\ref{sec_generalcomments} we present a general discussion of the
reduction of the 2.5D models of \citet{per97a,per00} to a series of 1D
models by neglecting radial terms.
We shall see that,
although this a rough approximation high above the disk,
it is quite accurate near the disk surface.
We reduce the 2.5D model of Pereyra and collaborators
to a series of 1D equations in \S\ref{sec_reduction}.
In \S\ref{sec_results} we present the results of our series of 1D models
for different disk radius.
We show that the critical point in the inner disk regions is found near
the disk surface,
allowing for an analysis of the solutions with steady wind mass loss
rates.
A summary and conclusions are presented in \S\ref{sec_sumcon}.

\section{General Comments}
\label{sec_generalcomments}

The analysis of existence for steady line-driven wind solutions
discussed in Paper~I and Paper~II correspond to 1D models.
Thus,
to apply them to a 2.5D model,
we first reduce the 2.5D model to a series of 1D models.
To do this we consider the following:
the wind in CVs tend to be vertical
(perpendicular to the disk).
There is observational,
theoretical,
and computational evidence for this.
In CVs the P-Cygni type line profiles are observed only in low
inclination systems
(close to disk face-on)
and with high inferred disk mass accretion rates
\citep[e.g.,][]{war95}.
This orientation angle effect was one of the observational properties
of CVs that led to the early suggestion by
\citet{cor82}
that the outflows detected in CVs originate from the accretion disk.

From a theoretical standpoint,
if the wind originates from a steady disk,
then the radial forces at wind base
(disk surface)
must be at equilibrium,
thus the wind must start off vertical.
We note here that by
``vertical''
we actually mean
``non-radial,''
since the wind starts off with an azimuthal component due to disk
rotation.
We also note here that although the disk wind must in principal start
vertical at its base,
this does not necessarily imply that the wind will continue
predominantly vertical.
For example,
in the case of QSOs/AGNs,
disk wind models
\citep{mur95,hil02}
indicate that the wind tends to quickly become radial in nature.

However,
computational 2.5D models of CV disk winds have produced winds that
tend to be vertical in nature.
Theoretical resonance line profiles calculated from these models are
consistent with observed line profiles in their general form,
in the magnitudes of wind velocities implied by the absorption
components,
in the FWHM of the emission components,
and in the strong dependence with inclination angle
\citep{per97a,per00}.

Therefore,
near the wind base
(disk surface),
a CV disk wind at a given radius can be well approximated by a vertical
1D model.
High above the disk rotational forces come into play
(no longer at equilibrium with the radial component of gravity),
streamlines collide,
and the density/velocity wind structure will be 2.5D
(i.e.,
 2D with an azimuthal velocity component)
\citep{per97a,per97b,per00,per03}.
Although a 1D vertical model
is not accurate high above the disk for the above reasons,
it does give accurate results near wind base.

As was shown by \citet{cas75} for early type stars,
and was discussed in Paper~II for 1D line-driven winds,
the wind mass loss rate is determined by the position of the critical
point.
As we show below,
for the CV parameters we use in this section
[taken from the models of
\citet{per00}],
at radii where there is significant contributions to the wind mass
loss,
the critical point is near the disk surface.
Therefore,
disk wind characteristics for 2.5D CV models can be addressed.

\section{Reduction to a Series of 1D Models}
\label{sec_reduction}

The equations of state, mass, and momentum of the 2.5D CV disk wind
model of
\citet{per00}
are, respectively:

\begin{equation}
P
=
\rho  \, b^2
\;\;\;\; ;
\end{equation}

\begin{equation}
\label{equ_pkb2_masscons}
{\partial\rho \over \partial t}
+ {1 \over r}{\partial(r \rho V_r) \over \partial r}
+ {\partial(\rho V_z) \over \partial z}
=
0
\;\;\;\; ;
\end{equation}

\noindent
and

\clearpage

\begin{eqnarray}
\rho{\partial V_r \over \partial t}
+ \, \rho V_r{\partial V_r \over \partial r}
- \rho {{V_\phi}^2 \over r}
+ \rho V_z{\partial V_r \over \partial z}
&=&
- \, \rho{GM_{wd} \over (r^2 + z^2)}{r \over (r^2+ z^2)^{1/2}}
- {\partial P \over \partial r} 
+ \rho{\kappa_e \, {\cal F}_r(r,z) \over c}
\nonumber
\\
&& \\
&& + \, \rho {\kappa_e \, S_r(r,z) \over c} \;
        \times
         k \left( \max \left[{ 1 \over \rho \kappa_e V_{th}}
                  \left|
                    {\partial V_z \over \partial z}
                  \right|\; ,\; 10^8 \right] \;
           \right)^\alpha
\;\;\;\; ;
\nonumber
\end{eqnarray}

\begin{equation}
\rho {\partial V_\phi \over \partial t}
+ \rho V_r{\partial V_\phi \over \partial r}
+ \rho {V_\phi V_r \over r}
+ \rho V_z{\partial V_\phi \over \partial z}
=
0
\;\;\;\; ;
\end{equation}

\begin{eqnarray}
\label{equ_pkb2_momz}
\rho{\partial V_z \over \partial t}
+ \rho V_r{\partial V_z \over \partial r}
+ \rho V_z{\partial V_z \over \partial z}
&=&
- \, \rho{GM_{wd} \over (r^2+z^2)}{z \over (r^2+z^2)^{1/2}}
- {\partial P \over \partial z}
+ \rho{\kappa_e \, {\cal F}_z(r,z) \over c}
\nonumber
\\
&& \\
&& + \, \rho {\kappa_e \, S_z(r,z) \over c} \;
     \times k \left( \max \left[{ 1 \over \rho \kappa_e V_{th}}
                 \left|
                   {\partial V_z \over \partial z}
                 \right|\; ,\; 10^8 \right] \;
              \right)^\alpha \nonumber
\;\;\;\; ;
\end{eqnarray}

\noindent
where
$P$
is the pressure,
$\rho$
is the density,
$b$
is the isothermal sound speed,
$V_r$,
$V_\phi$,
and
$V_z$
are the corresponding velocity components in cylindrical coordinates,
$G$
is the gravitational constant,
$M_{wd}$
is the mass of the white dwarf,
$\kappa_e$
is the Thomson cross section per mass,
$c$
is the speed of light,
$V_{th}$
is the ion thermal velocity,
and
$k$
and
$\alpha$
are the CAK line force parameters.
${\cal F}_z$
and
${\cal F}_r$
are the corresponding radiation flux components given by:

\begin{equation}
\label{equ_pkb2_fz}
{\cal F}_z(r,z)
=
\int_{r_i}^\infty \int_0^{2\pi} {Q(r') \over \pi} \;
   {z^2
    \over
    \left[\left(r^2+r'^2+z^2-2rr'\cos\phi\right)^{1/2}\right]^4 }
   \; r' d\phi dr'
\;\;\;\; ,
\end{equation}

\noindent
and

\begin{equation}
\label{equ_pkb2_fr}
{\cal F}_r(r,z)
=
\int_{r_i}^\infty \int_0^{2\pi} {Q(r') \over \pi} \;
  {z (r-r'\cos\phi)
   \over
   \left[\left(r^2+r'^2+z^2-2rr'\cos\phi\right)^{1/2}\right]^4 }
  \; r' d\phi dr'
\;\;\;\; ,
\end{equation}

\noindent
$S_z$
and
$S_r$
are defined as:

\begin{equation}
\label{equ_pkb2_sz}
S_z(r,z)
\equiv
\int_{r_i}^\infty \int_0^{2\pi} {Q(r') \over \pi} \;
  {z^{\alpha+2}
   \over
   \left[\left(r^2+r'^2+z^2-2rr'\cos\phi\right)^{1/2}\right]^{\alpha+4}}
  \; r' d\phi dr'
\;\;\;\; ,
\end{equation}

\noindent
and

\begin{equation}
\label{equ_pkb2_sr}
S_r(r,z)
\equiv
\int_{r_i}^\infty \int_0^{2\pi} {Q(r') \over \pi} \;
  {z^{\alpha+1}(r-r'\cos\phi)
   \over
   \left[\left(r^2+r'^2+z^2-2rr'\cos\phi\right)^{1/2}\right]^{\alpha+4}}
  \; r' d\phi dr'
\;\;\;\; ,
\end{equation}

\noindent
where in turn
$Q(r)$
is the rate of energy per area,
at a given radius
$r$,
radiated by a standard disk
\citep{sha73},
that is:

\begin{equation}
\label{equ_pkb2_q1}
Q(r)
=
{3 \dot M_{accr} G M_{wd} \over 8 \pi r^3}
  \left[1 - \left({r_i \over r} \right)^{1/2} \right]
\;\;\;\; ,
\end{equation}

\noindent
where
$r_i$
is the inner disk radius.
In CVs,
$r_i$ is approximately equal to the white dwarf radius.
We note that in equations~(\ref{equ_pkb2_fz})-(\ref{equ_pkb2_sr})
there is an implicit assumption that
$r_i \ll r_f$,
where
$r_f$
is the outer disk radius.

The total disk luminosity
${\cal L}_{disk}$
is given by

\begin{equation}
{\cal L}_{disk}
=
\int_{r_i}^{r_f} 4 \pi r Q(r) dr
\;\;\;\; ,
\end{equation}

\noindent
where
$r_f$
is the outer disk radius.
Thus

\begin{equation}
{\cal L}_{disk}
=
{\dot M_{accr} G M_{wd} \over 2 r_i}
  \left\{ 1 - {3r_i \over r_f}
    \left[ 1 - {2 \over 3}
      \left({r_i \over r_f} \right)^{1/2}
    \right]
  \right\}
\;\;\;\; ;
\end{equation}

\noindent
assuming
$r_i \ll r_f$

\begin{equation}
\label{equ_pkb2_ldisk}
{\cal L}_{disk}
\approx
{\dot M_{accr} G M_{wd} \over 2 r_i}
\;\;\;\; .
\end{equation}

We define

\begin{equation}
\label{equ_pkb2_gamma}
\Gamma
\equiv
{\kappa_e \, {\cal L}_{disk} \over 4 \pi G M_{wd} \, c}
\;\;\;\; .
\end{equation}

\noindent
Considering equations~(\ref{equ_pkb2_ldisk}) and (\ref{equ_pkb2_gamma}),
equation~(\ref{equ_pkb2_q1}) can be rewritten as

\begin{equation}
\label{equ_pkb2_q2}
Q(r)
=
3 \, {c \over \kappa_e} \, {G M_{wd} \over {r_i}^2} \, \Gamma 
  \left( r_i \over r \right)^3
  \left[1 - \left({r_i \over r} \right)^{1/2} \right]
\;\;\;\; .
\end{equation}

\noindent
Additionally,
we define two dimensionless functions
$\Upsilon(r,z)$
and
$\zeta(r,z)$:

\begin{equation}
\label{equ_pkb21_upsilon}
\Upsilon(r,z)
\equiv
\int_{r_i}^\infty \int_0^{2\pi} {1 \over \pi} \;
  \left( r_i \over r' \right)^3
  \left[1 - \left({r_i \over r'} \right)^{1/2} \right]
  {z^2
   \over
   \left[\left(r^2+r'^2+z^2-2rr'\cos\phi\right)^{1/2}\right]^4}
  \; r' d\phi dr'
\;\;\;\; ;
\end{equation}

\begin{equation}
\label{equ_pkb21_zeta}
\zeta(r,z)
\equiv
\int_{r_i}^\infty \int_0^{2\pi} {1 \over \pi} \;
  \left( r_i \over r' \right)^3
  \left[1 - \left({r_i \over r'} \right)^{1/2} \right]
  {z^{\alpha+2}
   \over
   \left[\left(r^2+r'^2+z^2-2rr'\cos\phi\right)^{1/2}\right]^{\alpha+4}}
  \; r' d\phi dr'
\;\;\;\; .
\end{equation}

\noindent
It follows that

\begin{equation}
{\cal F}_z(r,z)
=
3 \, {c \over \kappa_e} \, {G M_{wd} \over {r_i}^2} \, \Gamma \,
 \Upsilon(r,z)
=
3 {{\cal L}_{disk} \over 4 \pi {r_i}^2} \Upsilon(r,z)
\;\;\;\; ;
\end{equation}

\begin{equation}
S_z(r,z)
=
3 \, {c \over \kappa_e} \, {G M_{wd} \over {r_i}^2} \, \Gamma \,
 \zeta(r,z)
=
3 {{\cal L}_{disk} \over 4 \pi {r_i}^2} \zeta(r,z)
\;\;\;\; .
\end{equation}

In reducing the 2.5D model to a series of 1D models,
for simplicity,
we assume an isothermal wind.
We do not expect this approximation to significantly change the results
of this work,
as is illustrated in the intermediate models of
\citet{per00}
[see Figures~6 and 7 of
 \citet{per00}].
This, 
on one hand,
significantly simplifies the analysis,
and on the other hand gives a predefined temperature structure
[namely:
 $b^2(x) = {\rm constant}$]
that is a requirement of the 1D models of this work
(see Paper~I; Paper~II).
Thus the equation of state we assume,
for the application to the 2.5D CV disk wind model,
is

\begin{equation}
\label{equ_pkb2_state}
{P \over \rho}
=
b^2 \hskip 28pt ({\rm constant})\ .
\end{equation}

\noindent
We also assume that the streamlines,
starting from the disk at a given radius
$r$,
are predominantly vertical.
As discussed in \S\ref{sec_generalcomments},
this assumption is justified on observational,
theoretical,
and computational grounds.

The mass conservation equation is equivalent to the 1D models of
\citet{per97b},
namely,
for a given
$r$

\begin{equation}
\label{equ_pkb21_masscons}
\dot{M}
=
\rho \, V \, 4 \pi  R^2 \hskip 28pt (\rm constant)
\;\;\;\; ,
\end{equation}

\noindent
where we are determining the existence of 1D steady
(stationary)
solutions,
and where

\begin{equation}
R^2
\equiv
r^2+z^2
\;\;\;\; ,
\end{equation}

\noindent
and
$\dot{M}$
is an integration constant,
$z$
is the height above the disk,
and
$r$
is the radius where a given streamline begins.
Thus for each value of
$r$
considered,
a given 1D model is developed.
We note that in equation~(\ref{equ_pkb21_masscons}),
$\dot{M}$
is not the wind mass loss rate
(as in Paper~I and Paper~II),
but only an integration constant.
We have used the symbol
$\dot{M}$
following the notation of Paper~I and Paper~II.
However,
the wind mass loss rate
$\dot{M}_{wind}$
can be obtained by
(after first determining the values of
 $\dot{M}$
 for different values of
 $r$)

\begin{equation}
\dot{M}_{wind}
=
2 \int_{r_i}^{r_f} { \dot{M}(r) \over 4 \pi (r^2 + z_s^2) } \,
                   2 \pi r \, dr
\;\;\;\; .
\end{equation}

For
$z \ll r$,
$R^2 \approx r^2 \; (= \hbox{\rm constant for a given 1D model})$,
thus for
$z \ll r$,
$\rho V = {\rm constant}$
[equation~(\ref{equ_pkb21_masscons})]
that is the integration of equation~(\ref{equ_pkb2_masscons})
assuming stationary solutions and neglecting radial velocities.
Therefore,
equation~(\ref{equ_pkb21_masscons}) is accurate near the disk surface
and it is also accurate towards infinity
($z \gg r$,
 $R^2 \approx z^2$).
However,
equation~(\ref{equ_pkb21_masscons}) is not accurate at intermediate
values of
$z$
where 2.5D effects become important
\citep{per97a,per97b,per00}.
Using the accurate results near the disk surface,
as discussed in \S\ref{sec_generalcomments},
we study whether wind solutions with steady mass loss rates exist.

In reducing the 2.5D model to a series of 1D models,
we additionally neglect the effect of a maximum possible value for the
line radiation force
\citep{cas75,abb82},
that are represented in the models of
\citet{per97a} and \citet{per00}
by including a
``$\max$''
function in the line force terms
[see equation~(\ref{equ_pkb2_momz})].
On one hand,
we do not expect this approximation to significantly change the results
of this work.
In the intermediate models of
\citet{per00}
this effect was found to be most significant in the low density region
high above the inner disk region.
On the other hand,
this allows the momentum equation to take a form equivalent to
the momentum equation of Paper~II,
that is the form of the momentum equation assumed in the
1D models of this work.

Thus,
within the assumption of a predominately vertical wind,
assuming stationary solutions,
neglecting radial velocities,
and considering equation~(\ref{equ_pkb2_momz}),
the momentum equation for the 1D models
(at a given
 $r$)
becomes

\newpage

\begin{eqnarray}
\rho V {d V \over d z}
& = &
- \, \rho{GM_{wd} \over R^2} \, {z \over R} \,
+ \, \rho { 3 \, GM_{wd} \over r_i^2} \, \Gamma \, \Upsilon(r,z) \,
- \, {dP \over dz} \,
\nonumber
\\
&&
\\
&& + \, \rho { 3 \, GM_{wd} \over {r_i}^2} \, \Gamma \, \zeta(r,z) \,
   k \left( { 1 \over \rho \kappa_e V_{th}}
         {\partial V \over \partial z} \right)^\alpha
\;\;\;\; ,
\nonumber
\end{eqnarray}

\noindent
where
$\Upsilon(r,z)$
and
$\zeta(r,z)$
are given by equations~(\ref{equ_pkb21_upsilon}) and
(\ref{equ_pkb21_zeta}),
respectively.
The series of 1D models derived from the 2.5D disk wind model,
can be represented within the 1D models of Paper~II,
through the following parameterization:

\begin{eqnarray}
  B(z)
&=&
  {G M_{wd} \over R^2} \, {z \over R}
- {3 G M_{wd} \over r_i^2} \Gamma \Upsilon(r,z)
\;\;\;\; ,
\\
& & \nonumber
\\
  A(z)
&=&
  4 \, \pi \, R^2 = 4 \, \pi \, \left(r^2 + z^2 \right)
\;\;\;\; ,
\\
& & \nonumber \\
  \hbox{and \ \ \ \ \ \ \ \ } \gamma(z)
&=&
  {3 \, G M_{wd} \over {r_i}^2} \, \Gamma \, \, \zeta(r,z) \,
      k \left(1 \over \kappa_e V_{th}\right)^\alpha
\;\;\;\; ,
\end{eqnarray}

\noindent
where the body force $B(z)$,
the area function
$A(z)$,
and the line opacity weighted flux
$\gamma(z)$
are defined for a given $r$.
Thus,
the mass conservation equation and the momentum equation have the form
of the equations of Paper~II.
The isothermal equation of motion is therefore,

\begin{equation}
  \left(1 -{b^2 \over V^2} \right) A V {dV \over dz}
=
- \, B A
+ \, \gamma A \left({A \over \dot{M}} V { dV \over dz } \right)^\alpha
+ b^2  {dA \over dz}
\;\;\;\; .
\end{equation}

Defining the characteristic distance
$r_0$,
gravitational force
$B_0$,
area $A_0$,
and line-opacity weighted flux
$\gamma_0$
(see Paper~I; Paper~II)
through

\begin{equation}
  r_0
=
  r
\hskip 16pt
;
\hskip 16pt
  B_0
=
  {G M_{wd} \over r^2}
\hskip 16pt
;
\hskip 16pt
  A_0
= 
  4 \pi r^2
\hskip 16pt
;
\hskip 16pt
  \gamma_0
=
  {3 \, G M_{wd} \over r_i^2} \, \Gamma \,
      k \left(1 \over \kappa_e V_{th}\right)^\alpha
\;\;\;\; ;
\end{equation}

\noindent
the equation of motion becomes

\begin{equation}
  \left(1-{s \over \omega} \right) a \, {d \omega \over d x}
=
- g a
+ f a \left({a \over \dot{m}} {d\omega \over dx} \right)^\alpha
+ 4 s x
\;\;\;\; ,
\end{equation}

\noindent
where

\begin{equation}
  x
=
  {z \over r}
\;\;\;\; ;
\end{equation}

\begin{equation}
  g
=
  {x \over \left(1+x^2\right)^{3/2}}
-
  {3 \Gamma \over x_i^2} \, \Upsilon
\hskip 24pt
;
\hskip 24pt
  x_i
\equiv
  {r_i \over r}
\;\;\;\; ;
\end{equation}

\begin{equation}
  a
=
  1 + x^2
\;\;\;\; ;
\end{equation}

\begin{equation}
  f
= 
  {1 \over \alpha^\alpha (1-\alpha)^{1-\alpha}} \, \zeta
\;\;\;\; .
\end{equation}

\noindent
The parameters
$s$, $\omega$, and $\dot{m}$
(see Paper~I; Paper~II)
are given by

\begin{equation}
  s
\equiv
  {b^2 \over 2 W_0}
\hskip 24pt
;
\hskip 24pt
  \omega
\equiv
  {W \over W_0}
\hskip 24pt
;
\hskip 24pt
  \dot{m}
\equiv
  {\dot{M} \over \dot{M}_{CAK}}
\;\;\;\; ,
\end{equation}

\noindent
where

\begin{equation}
  W_0
\equiv
  B_0 r_0
\hskip 24pt
;
\hskip 24pt
  W
\equiv
  {V^2 \over 2}
\hskip 24pt
;
\hskip 24pt
  \dot{M}_{CAK}
\equiv
  \alpha (1-\alpha)^{(1 - \alpha) / \alpha}
    { \left( \gamma_0 A_0 \right)^{1 / \alpha}
      \over
      \left(B_0 A_0 \right)^{(1-\alpha) / \alpha}
    }
\;\;\;\; .
\end{equation}

Taking
$x_0=0$
and
$q_0=0$,
the spatial variable
$q$
(see Paper~II)

\begin{equation}
\label{equ_q}
  q
\equiv
  \int\displaylimits_{x_0}^{x} {1 \over a(x')} \, dx' + q_0
\;\;\;\; ,
\end{equation}

\noindent
becomes

\begin{equation}
  q
=
  \arctan(x)
\;\;\;\; .
\end{equation}

\noindent
The equation of motion,
expressed in terms of $q$,
becomes

\begin{equation}
  \left(1-{s \over \omega} \right) {d \omega \over dq}
=
  h(q)
+ f a \left({1 \over \dot{m}} {d\omega \over dq} \right)^\alpha
\;\;\;\; ,
\end{equation}

\noindent
where the function
$h(q)$
is now given by

\begin{equation}
  h(q)
=
- g a
+ 4 s \tan(q)
\;\;\;\; .
\end{equation}

The nozzle function $n$ for each $r$ is given by (Paper~II):

\begin{eqnarray}
\label{equ_n}
  n(q)
\equiv
&&
  \alpha (1-\alpha)^{(1 - \alpha) / \alpha}
    {(fa)^{1 / \alpha} \over
     (-h)^{(1-\alpha) / \alpha}}
\hskip 48pt
{\rm for}
\hskip 24pt
  h(q)
<
  0
\nonumber \\
&&
\left[
=
  \alpha (1-\alpha)^{(1 - \alpha) / \alpha}
    {(fa)^{1 / \alpha} \over
     ( g a - 4 s \tan(q) )^{(1-\alpha) / \alpha}
    }
\right]
\;\;\;\; .
\end{eqnarray}

\noindent
Additionally we shall recall the definition of the $\beta$ function
as defined in Paper~I and Paper~II

\begin{equation}
\label{equ_beta}
  \beta(\omega)
\equiv
  1 - {s \over w}
\;\;\;\; ,
\end{equation}
that we shall apply in the following section.

Thus,
we have reduced the 2.5D models to a series of 1D models by solving the
hydrodynamic equations in the vertical direction and neglecting radial
terms.
That is,
for a given radial distance
$r$,
we have a 1D model with
$q=\arctan(x)=\arctan(z/r)$
as the independent spatial variable.
As discussed above these models are accurate near the disk surface.
Since,
as we see below,
the corresponding critical point for inner radii is near the disk
surface,
the 1D models for the inner disk regions determine that disk wind flows
with steady mass loss rates are possible.
The models can be used to estimate local mass loss rates
(by determining local density at the sonic point)
as well as velocity and density wind structure near the disk surface
in the inner region.

\section{Results}
\label{sec_results}

For the 1D wind models,
we implement the set of parameters used by
\citet{per00}
corresponding to a CV,
namely:

\begin{eqnarray}
M_{wd}
=
0.6 M_{\sun} 
\hskip 14pt
;
\hskip 14pt
{\cal L}_{disk}
&=&
{\cal L}_\sun
\hskip 33pt
;
\hskip 33pt
z_s = 0.0229 R_{\sun} 
\;\;\;\; ;
\nonumber
\\
\\
r_i
=
0.01 R_{\sun}
\hskip 21pt
;
\hskip 21pt
b
&=&
10 \; \hbox{km s}^{-1}
\hskip 14pt
;
\hskip 14pt
V_{th} = 2.67 \; \hbox{km s}^{-1}
\;\;\;\; .
\nonumber
\end{eqnarray}

\noindent
We use the line force parameters also used by
\citet{per00}
[and
 \citet{cas75}
 and
 \citet{abb82}],
namely

\begin{equation}
k
=
1/3 
\hskip 14pt
;
\hskip 14pt
\alpha
=
0.7 \ .
\end{equation}

As with the analysis of the isothermal CAK stellar wind
(Paper~II),
we first consider,
for a given
$r$,
the
$h$
function for the 1D vertical wind model with the flux distribution of a
standard
\citet{sha73}
disk.
As discussed in Paper~II,
$h$
must be negative at the critical point
(i.e.,
 $h$ must be negative in order for there to exist values of $q$,
 $\omega$,
 and
 $d\omega/dq$
 that satisfy the critical point conditions).
Second,
we consider the nozzle function,
that as we found in Paper~II,
for the case of an isothermal wind must be a locally increasing
function at the critical point
(i.e.,
 it must have a positive spatial derivative at the critical point).
Third,
for the spatial points where the first two conditions hold,
through the nozzle function
$n$,
the
$\beta$
function,
the normalized wind mass loss rate
$\dot{m}$,
and the calculation of
$d^2\omega/dq^2$
under critical point conditions,
as discussed in Paper~II,
we determine if a local solution exists by verifying whether or not
the condition

\begin{equation}
\label{equ_local4}
\beta''(\omega) \, \dot{m} \, (\omega')^2
+ \beta'(\omega) \, \dot{m} \, \omega''
- n''(q)
<
  0
\;\;\;\; 
\end{equation}

\noindent
holds.
Fourth,
within the set of points that fulfill the first three conditions,
we iteratively determine the position of the critical point such that
the wind reaches sound speed at the sonic height of the model
(as discussed in Paper~II).

As a consistency check,
for a given
$r$,
we superimpose a plot of the nozzle function
$n$
and the
$\beta(\omega) \, \dot{m}$
function evaluated with the corresponding velocities
$V(z)$.
As discussed in Paper~II,
in the supersonic wind regime where the
$h$
function is negative
(Region II),
for points other than the critical point,
the following conditions must hold

\begin{equation}
  \beta(\omega) \, \dot{m}
<
  n(q)
\hskip 24pt
({\rm for \ } \omega > s {\rm \ and \ }
              h(q) < 0 {\rm \ and \ } q \not= q_c)
\;\;\;\; ,
\label{equ_app1}
\end{equation}

\noindent
and at the critical point

\begin{equation}
  \beta(\omega_c) \, \dot{m}
=
  n(q_c)
\;\;\;\; .
\label{equ_app2}
\end{equation}

\begin{figure}
\epsscale{1.0}
\plotone{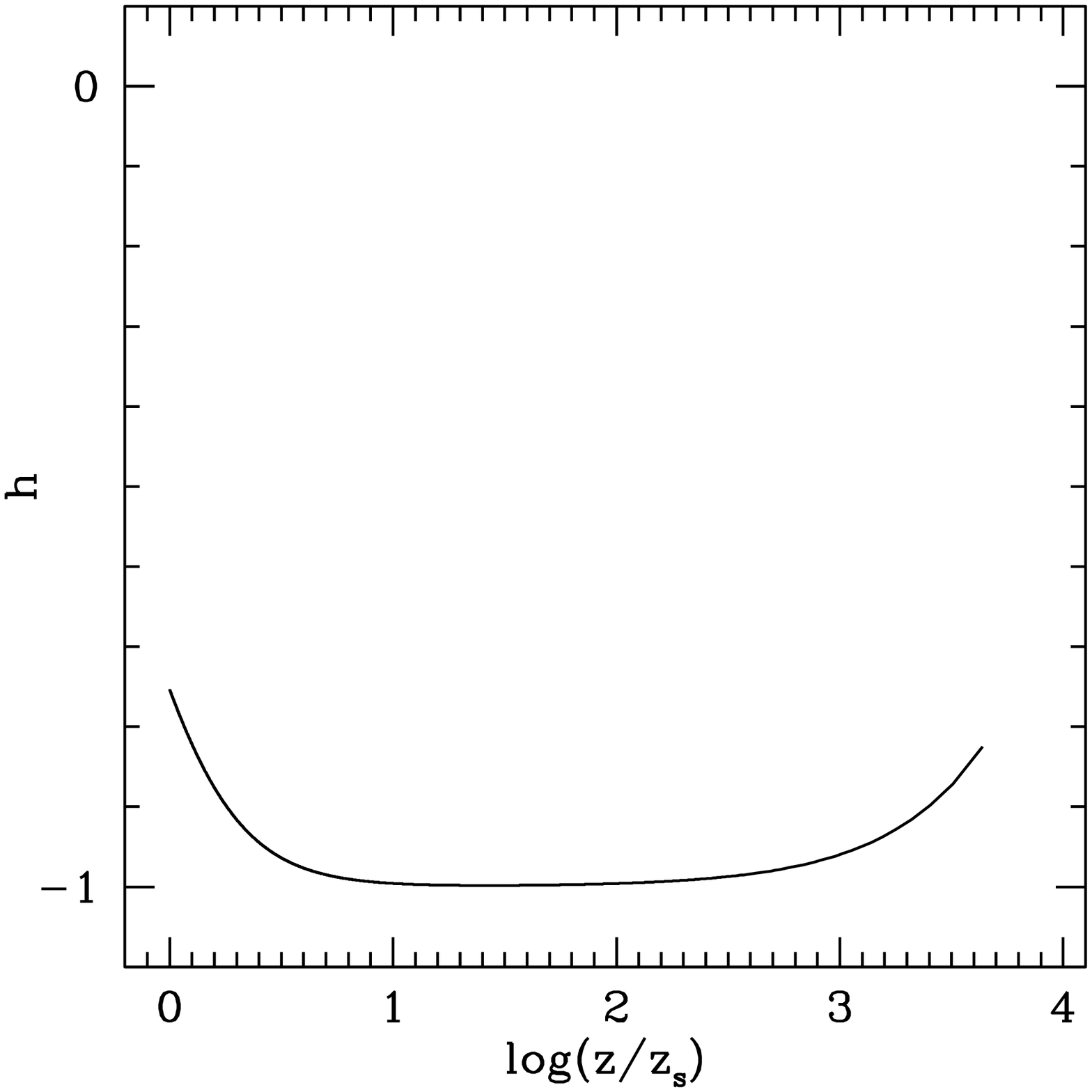}
\caption{
The
$h$
function for the 1D vertical wind model with the flux distribution of a
standard Shakura-Sunyaev disk at
$r=2 \, r_i$.
A necessary requirement for the critical point is that the $h$ function
must be negative at that point.
This figure shows that all points ranging from the sonic point to beyond
1000 times the sonic height fulfill this condition.
The physical parameters used are:
$M = 0.6 \, M_\sun$,
${\cal L}_{disk} = {\cal L}_\sun$,
$z_s = 0.0229 \,R_\sun$,
$r_i=0.01R_\sun$,
$b = 10 \, {\rm km \, s}^{-1}$,
$V_{th} = 2.67 \, {\rm km \, s}^{-1}$,
$k = 1/3$,
and
$\alpha = 0.7$~.
}
\label{fig_pkb21_02_h}
\end{figure}

\begin{figure}
\epsscale{1.0}
\plotone{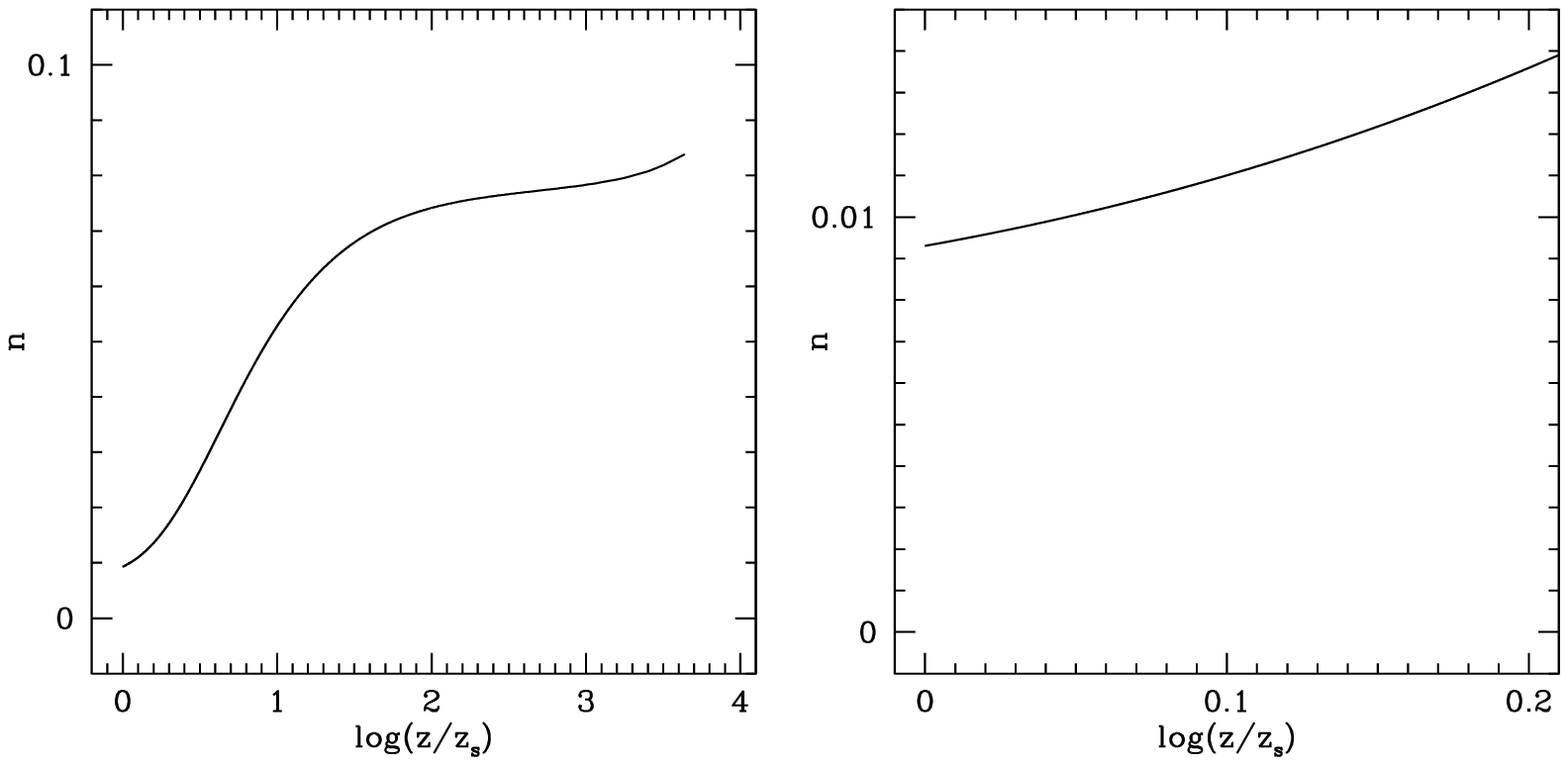}
\caption{
The nozzle function $n$ for the 1D vertical wind model with the flux
distribution of a standard Shakura-Sunyaev disk at
$r=2 \, r_i$.
A necessary requirement for the critical point in an isothermal
line-driven wind is that the nozzle function must be locally increasing 
(i.e.,
 $dn/dq > 0$
 [$dn/dz > 0$])
at that point.
This figure shows that all points ranging from the sonic point to beyond
1000 times the sonic height fulfill this condition.
The physical parameters used are the same as in
Figure~\ref{fig_pkb21_02_h}.
}
\label{fig_pkb21_02_n}
\end{figure}

\begin{figure}
\epsscale{0.95}
\plotone{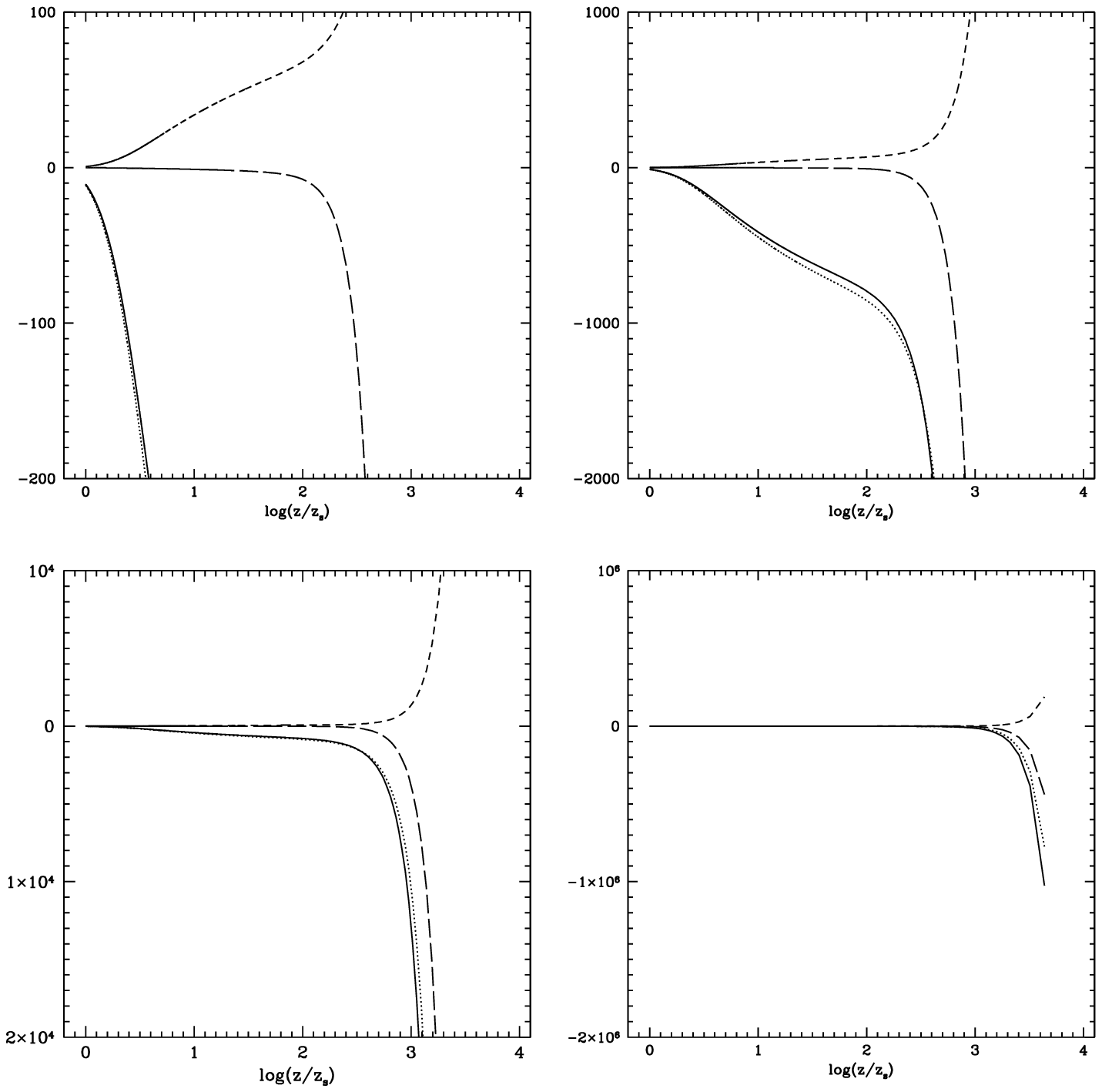}
\caption{
Local solution existence for the 1D vertical wind model with the flux
distribution of a standard Shakura-Sunyaev disk at
$r=2 \, r_i$:
a local solution exists in the vicinity of a critical point
$z$
if the expression
$\beta''\, \dot{m} \, (\omega')^2 + \beta' \, \dot{m} \, \omega'' - n''$
(solid curve)
is negative.
This figure shows that all points ranging from the sonic point to beyond
1000 times the sonic height fulfill this condition.
In this plot the terms
$\beta'' \, \dot{m} \, (\omega')^2$
(dotted curve),
$\beta' \, \dot{m} \, \omega''$
(short dash curve),
and
$-n''$
(long dash curve)
are also shown.
The physical parameters used are the same as in
Figure~\ref{fig_pkb21_02_h}.
}
\label{fig_pkb21_02_l}
\end{figure}

\begin{figure}
\epsscale{1.0}
\plotone{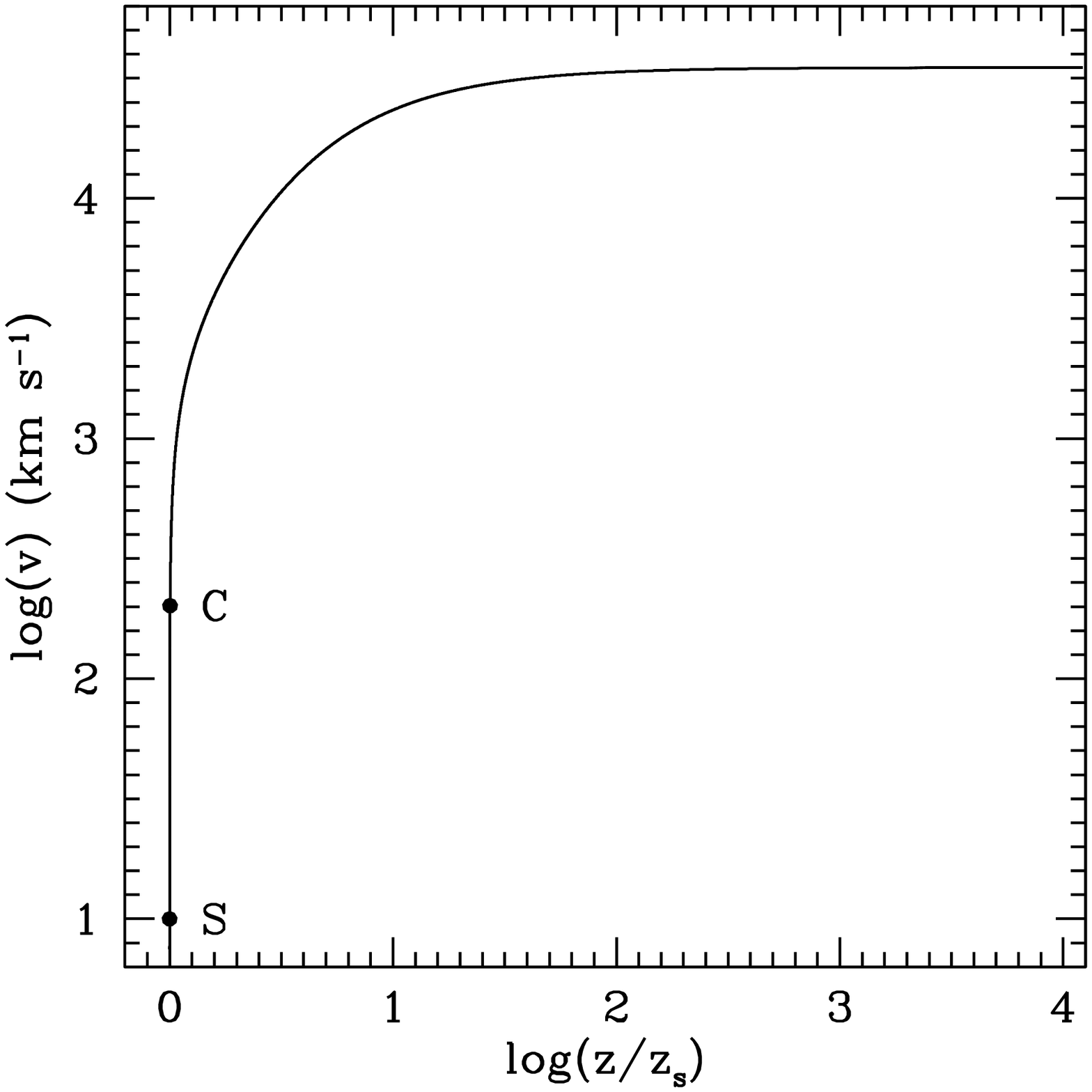}
\caption{
Velocity vs. position for the 1D vertical wind model with the flux
distribution of a standard Shakura-Sunyaev disk at
$r=2 \, r_i$.
The critical point position is determined with the condition that,
upon integration of the equation of motion,
the correct sonic point position is found.
``C''
indicates the critical point and
``S''
indicates the sonic point.
The physical parameters used are the same as in
Figure~\ref{fig_pkb21_02_h}.
}
\label{fig_pkb21_02_v}
\end{figure}

\begin{figure}
\epsscale{0.95}
\plotone{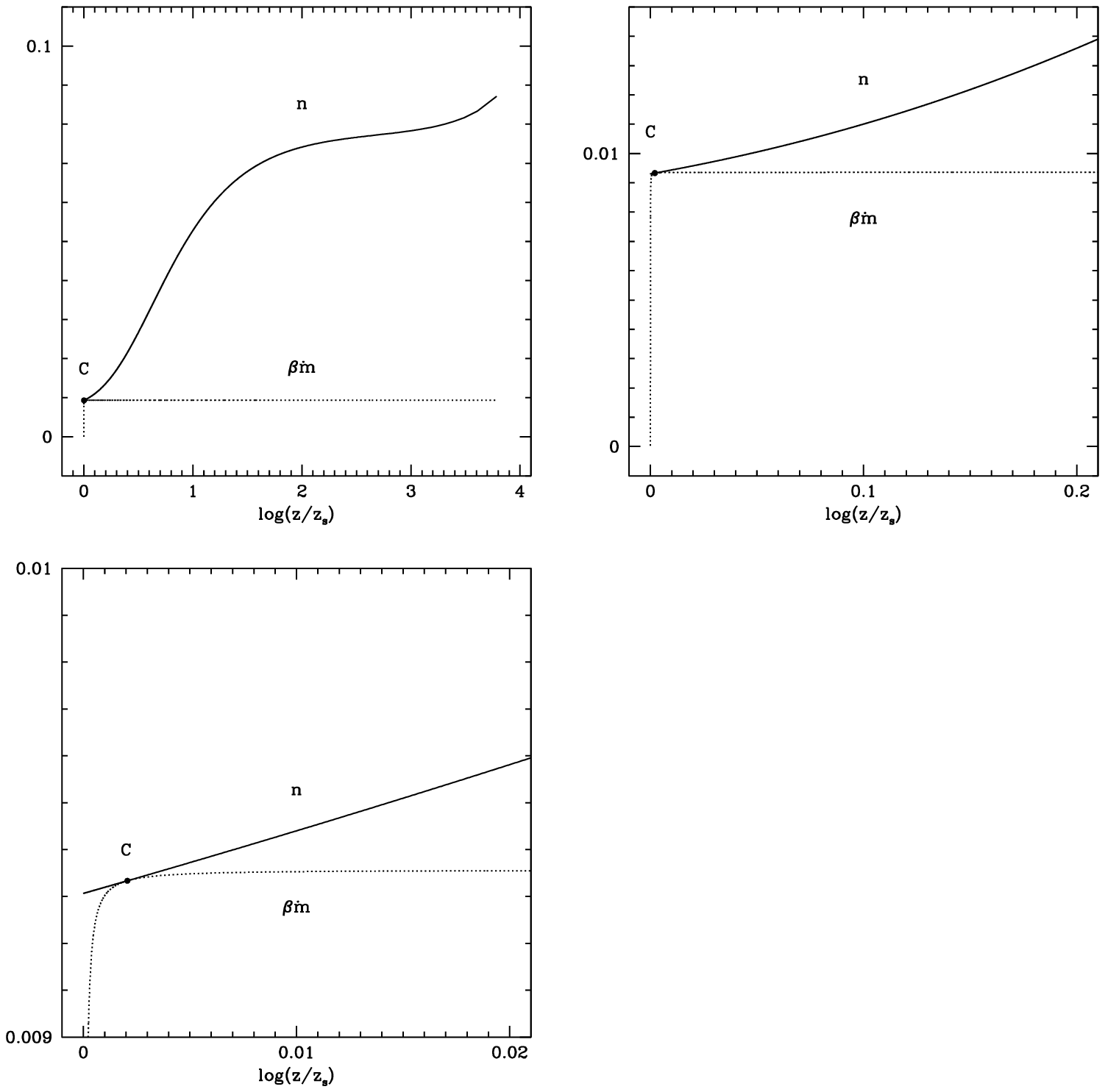}
\caption{
Necessary condition for the global solution existence for the
1D vertical wind model with flux distribution of a standard
Shakura-Sunyaev disk at
$r=2 \, r_i$
for the critical point shown in Figure~\ref{fig_pkb21_02_v}:
upon the integration of the equation of motion,
it must hold that
$\beta(\omega) \, \dot{m} < n(q)$
at points other than the critical point
[when
 $h(q)<0$
 and the wind is supersonic],
and
$\beta(\omega) \, \dot{m} = n(q)$
at the critical point.
Presented here is the nozzle function
$n$
(solid curve)
and the
$\beta \, \dot{m}$
function
(dotted curve)
vs. position.
``C''
indicates the critical point.
The physical parameters used are the same as in
Figure~\ref{fig_pkb21_02_h}.
}
\label{fig_pkb21_02_nb}
\end{figure}

For
$r = 2 \, r_i$,
Figure~\ref{fig_pkb21_02_h} shows that the
$h$
function is negative
from the sonic point to beyond 1000 times the sonic height.
Figure~\ref{fig_pkb21_02_n} shows that the nozzle function
$n$,
corresponding to a vertical streamline at
$r = 2 \, r_i$,
is a monotonically increasing function from the sonic point to beyond
1000 times the sonic height.
It is shown,
in Figure~\ref{fig_pkb21_02_l},
that equation~(\ref{equ_local4}) holds from the sonic height
to beyond 1000 times the sonic height,
showing that local solutions to the equation of motion exist throughout 
the fore mentioned spatial region.
By
``local solution''
at a given spatial point,
we mean the integration of the equation of motion in the vicinity
of the given point,
assuming the point to be the critical point.
Figure~\ref{fig_pkb21_02_v} shows the velocity vs. height obtained
upon integrating the equation of motion with the condition of
achieving the correct sonic height of the model.
As a consistency check,
Figure~\ref{fig_pkb21_02_nb} verifies that equations~(\ref{equ_app1})
and (\ref{equ_app2}) hold for the solution obtained.

\begin{figure}
\epsscale{1.0}
\plotone{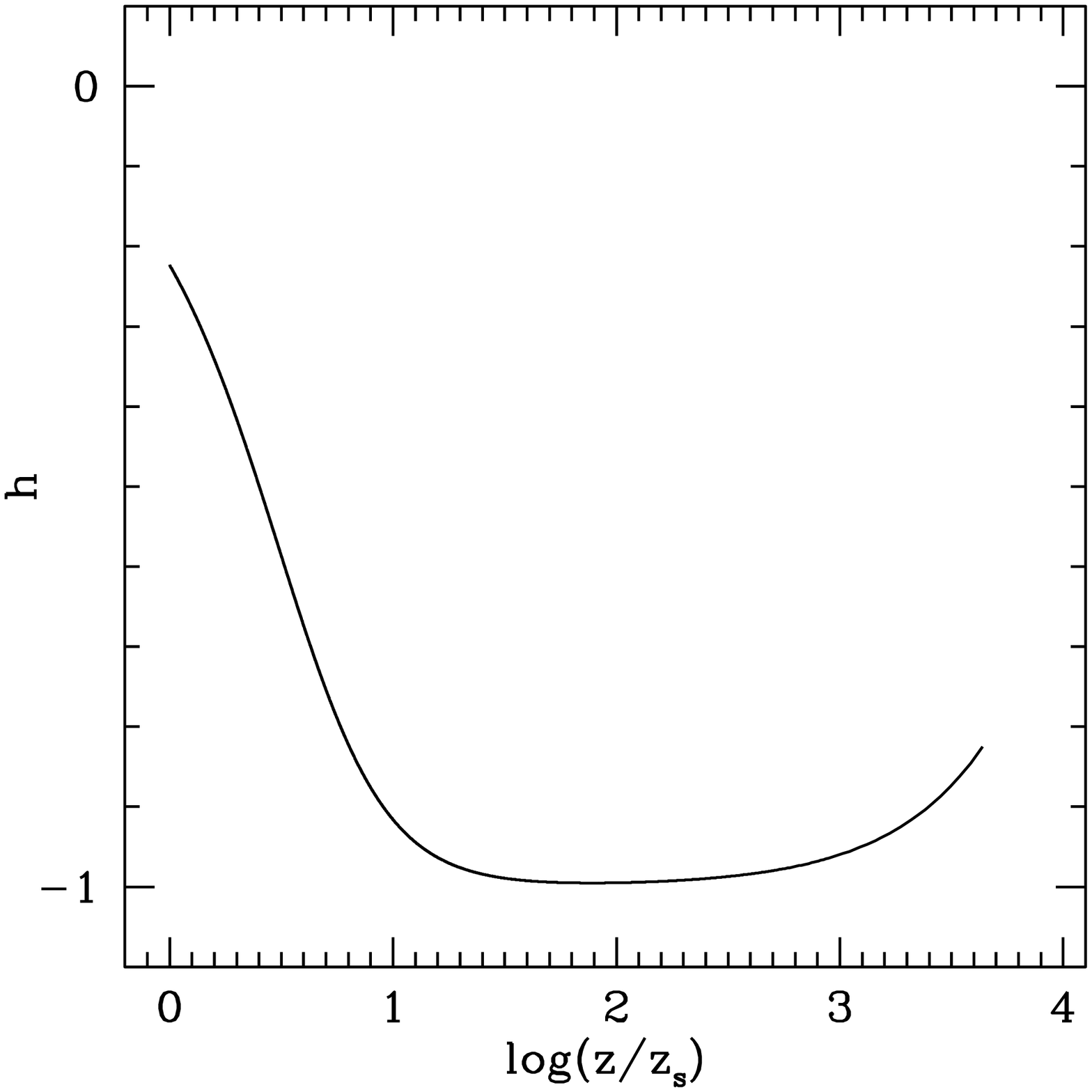}
\caption{
The
$h$
function for the 1D vertical wind model with the flux distribution of a
standard Shakura-Sunyaev disk at
$r=10 \, r_i$.
A necessary requirement for the critical point is that the
$h$
function must be negative at that point.
This figure shows that all points ranging from the sonic point to beyond
1000 times the sonic height fulfill this condition.
The physical parameters used are the same as in
Figure~\ref{fig_pkb21_02_h}.
}
\label{fig_pkb21_10_h}
\end{figure}

\begin{figure}
\epsscale{1.0}
\plotone{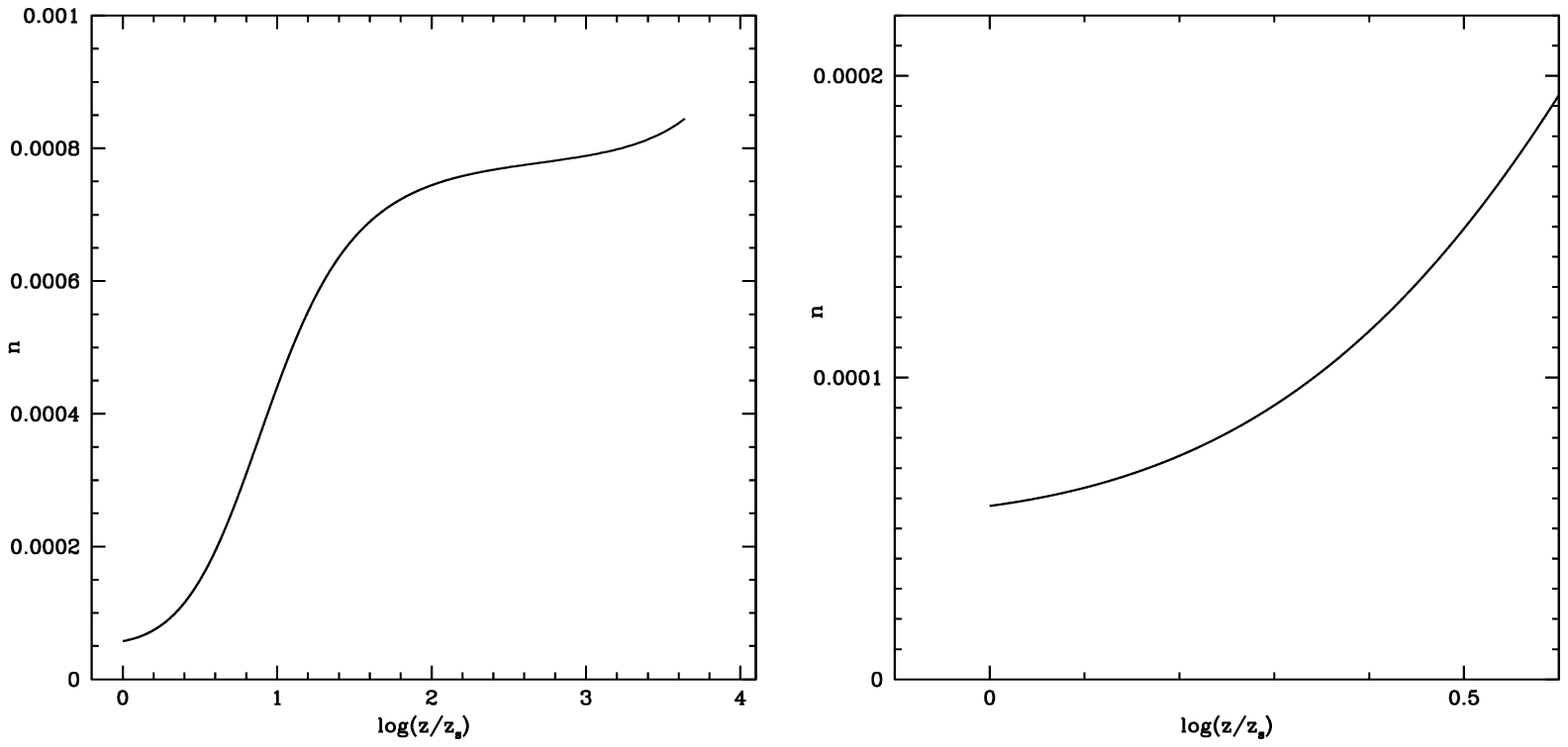}
\caption{
The nozzle function
$n$
for the 1D vertical wind model with the flux distribution of a standard
Shakura-Sunyaev disk at
$r=10 \, r_i$.
A necessary requirement for the critical point in an isothermal
line-driven wind is that the nozzle function must be locally increasing 
(i.e.,
 $dn/dq > 0$)
at that point.
This figure shows that all points ranging from the sonic point to beyond
1000 times the sonic height fulfill this condition.
The physical parameters used are the same as in
Figure~\ref{fig_pkb21_02_h}.
}
\label{fig_pkb21_10_n}
\end{figure}

\begin{figure}
\epsscale{0.95}
\plotone{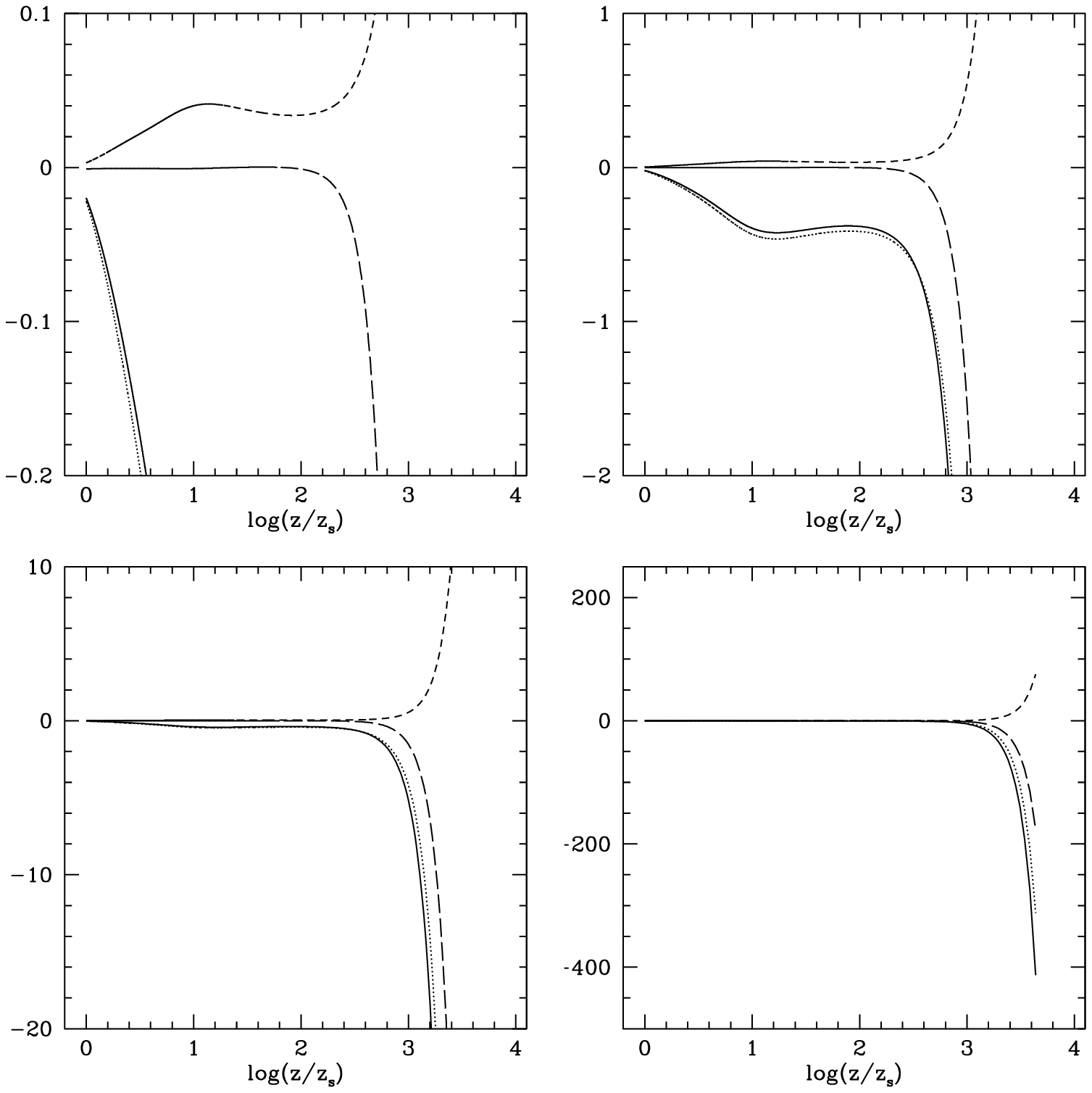}
\caption{
Local solution existence for the 1D vertical wind model with the flux
distribution of a standard Shakura-Sunyaev disk at
$r=10 \, r_i$:
a local solution exists in the vicinity of a critical point
$z$
if the expression
$\beta''\, \dot{m} \, (\omega')^2 + \beta' \, \dot{m} \, \omega'' - n''$
(solid curve)
is negative.
This figure shows that all points ranging from the sonic point to beyond
1000 times the sonic height fulfill this condition.
In this plot the terms
$\beta'' \, \dot{m} \, (\omega')^2$
(dotted curve),
$\beta' \, \dot{m} \, \omega''$
(short dash curve),
and
$-n''$
(long dash curve)
are also shown.
The physical parameters used are the same as in
Figure~\ref{fig_pkb21_02_h}.
}
\label{fig_pkb21_10_l}
\end{figure}

\begin{figure}
\epsscale{0.95}
\plotone{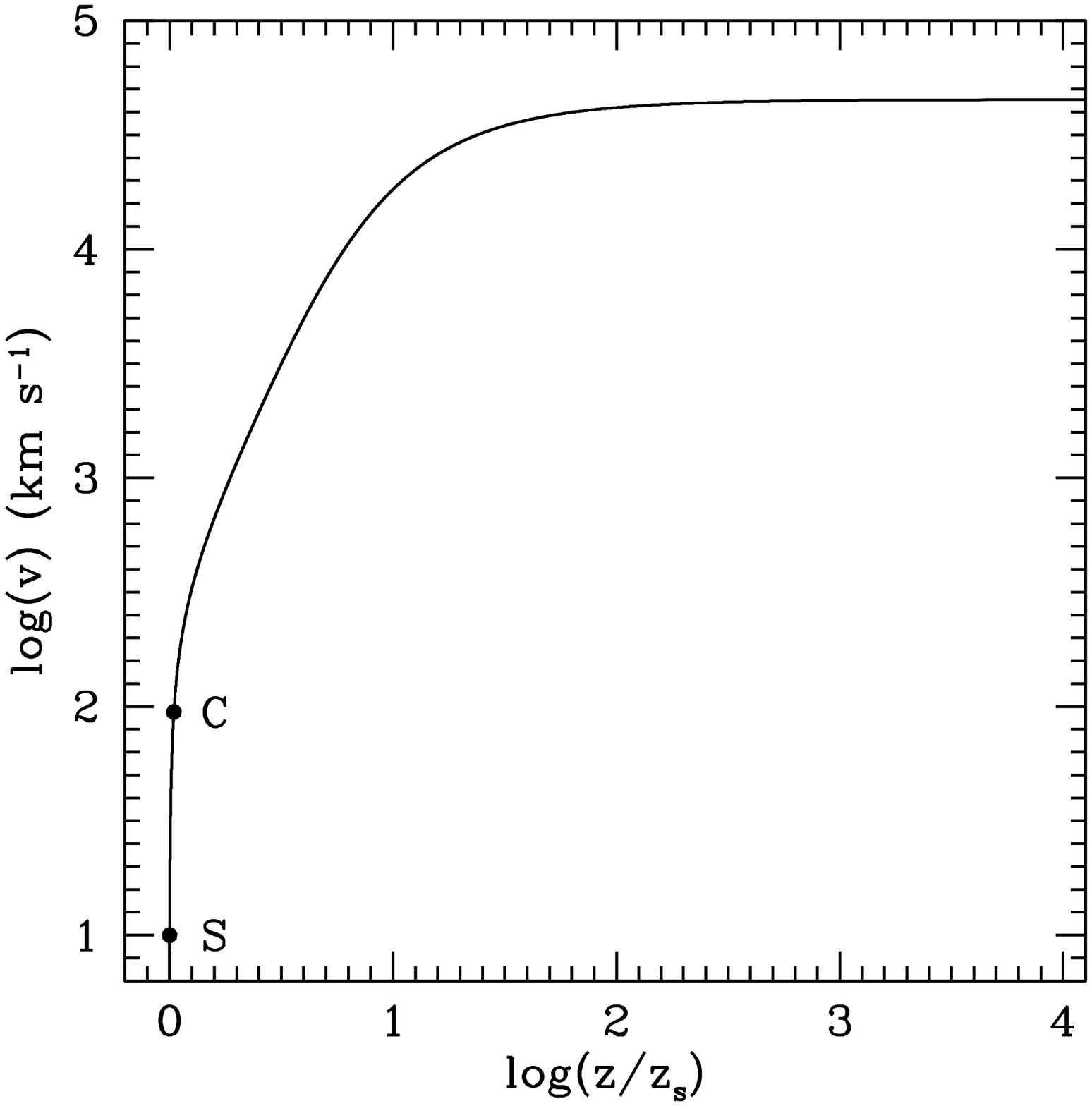}
\caption{
Velocity vs. position for the 1D vertical wind model with the flux
distribution of a standard Shakura-Sunyaev disk at
$r=10 \, r_i$.
The critical point position is determined with the condition that,
upon integration of the equation of motion,
the correct sonic point position is found.
``C''
indicates the critical point and
``S''
indicates the sonic point.
The physical parameters used are the same as in
Figure~\ref{fig_pkb21_02_h}.
}
\label{fig_pkb21_10_v}
\end{figure}

\begin{figure}
\epsscale{0.95}
\plotone{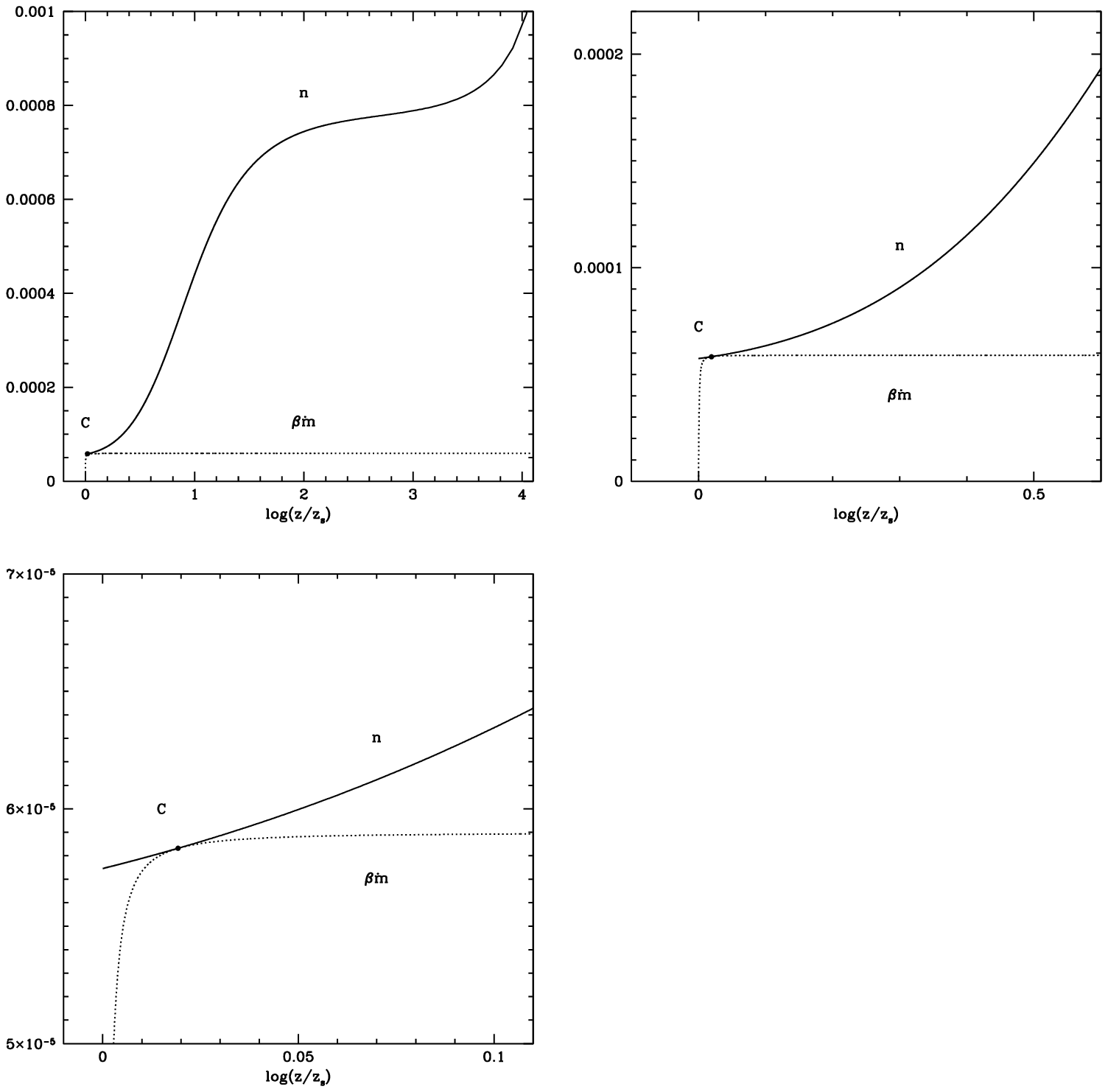}
\caption{
Necessary condition for the global solution existence for the
1D vertical wind model with the flux distribution of a standard
Shakura-Sunyaev disk at
$r=10 \, r_i$
for the critical point shown in Figure~\ref{fig_pkb21_10_v}:
upon the integration of the equation of motion,
it must hold that
$\beta(\omega) \, \dot{m} < n(q)$
at points other than the critical point
[when
 $h(q)<0$
 and the wind is supersonic],
and
$\beta(\omega) \, \dot{m} = n(q)$
at the critical point.
Presented here is the nozzle function
$n$
(solid curve)
and the
$\beta \, \dot{m}$
function
(dotted curve)
vs. position.
``C''
indicates the critical point.
The physical parameters used are the same as in
Figure~\ref{fig_pkb21_02_h}.
}
\label{fig_pkb21_10_nb}
\end{figure}

\begin{figure}
\epsscale{1.0}
\plotone{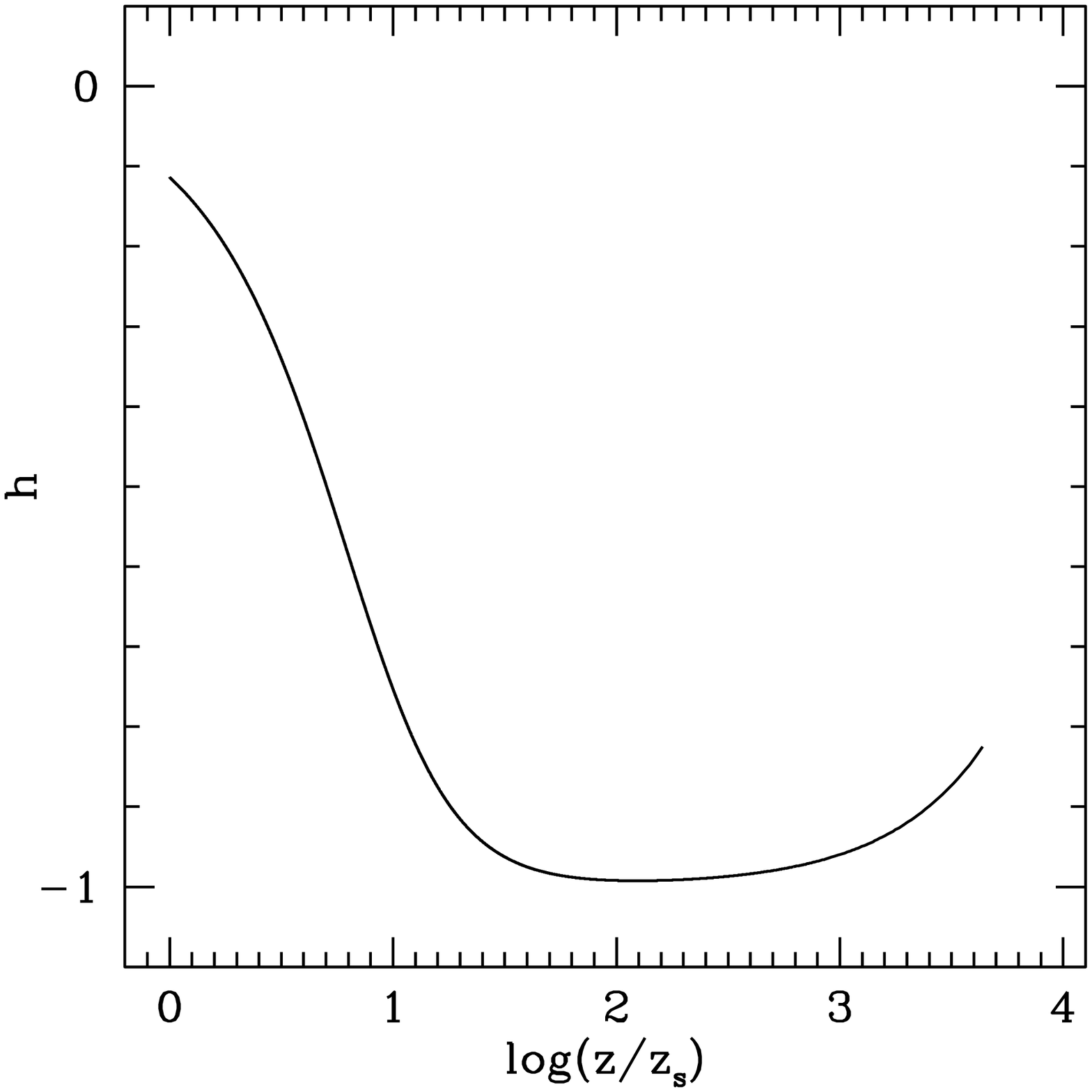}
\caption{
The
$h$
function for the 1D vertical wind model with the flux
distribution of a standard Shakura-Sunyaev disk at
$r=20 \, r_i$.
A necessary requirement for the critical point is that the
$h$
function
must be negative at that point.
This figure shows that all points ranging from the sonic point to beyond
1000 times the sonic height fulfill this condition.
The physical parameters used are the same as in
Figure~\ref{fig_pkb21_02_h}.
}
\label{fig_pkb21_20_h}
\end{figure}

\begin{figure}
\epsscale{1.0}
\plotone{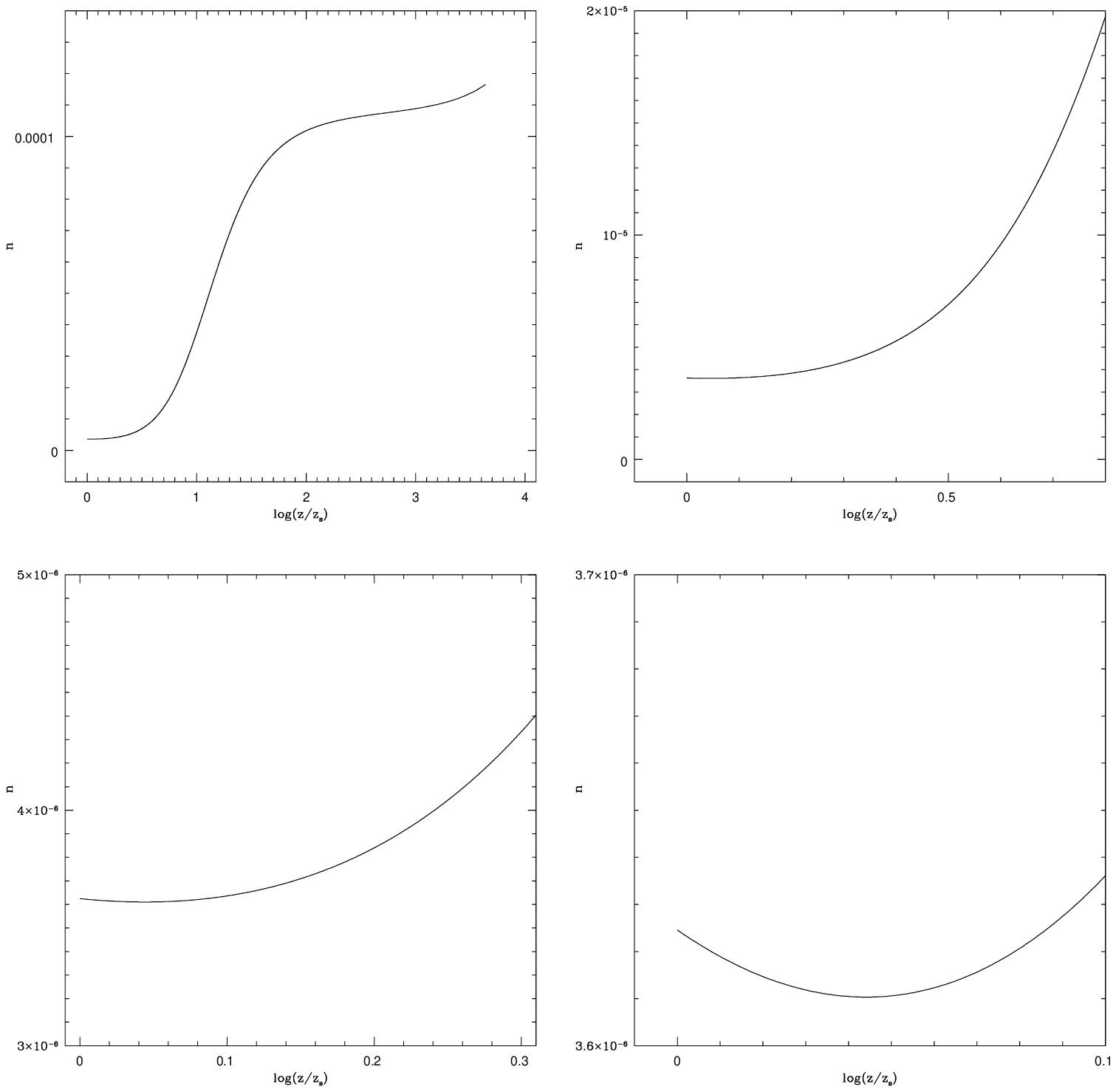}
\caption{
The nozzle function
$n$
for the 1D vertical wind model with the flux
distribution of a standard Shakura-Sunyaev disk at
$r=20 \, r_i$.
A necessary requirement for the critical point in an isothermal
line-driven wind is that the nozzle function must be locally increasing 
(i.e.,
 $dn/dq > 0$)
at that point.
This figure shows that all points ranging from
$\approx 1.107$
times the sonic height to beyond 1000 times the sonic height fulfill
this condition.
However,
it is also shown that points ranging from the sonic point up to
$\approx 1.107$
times the sonic height do not fulfill the condition.
The physical parameters used are the same as in
Figure~\ref{fig_pkb21_02_h}.
}
\label{fig_pkb21_20_n}
\end{figure}

\begin{figure}
\epsscale{0.95}
\plotone{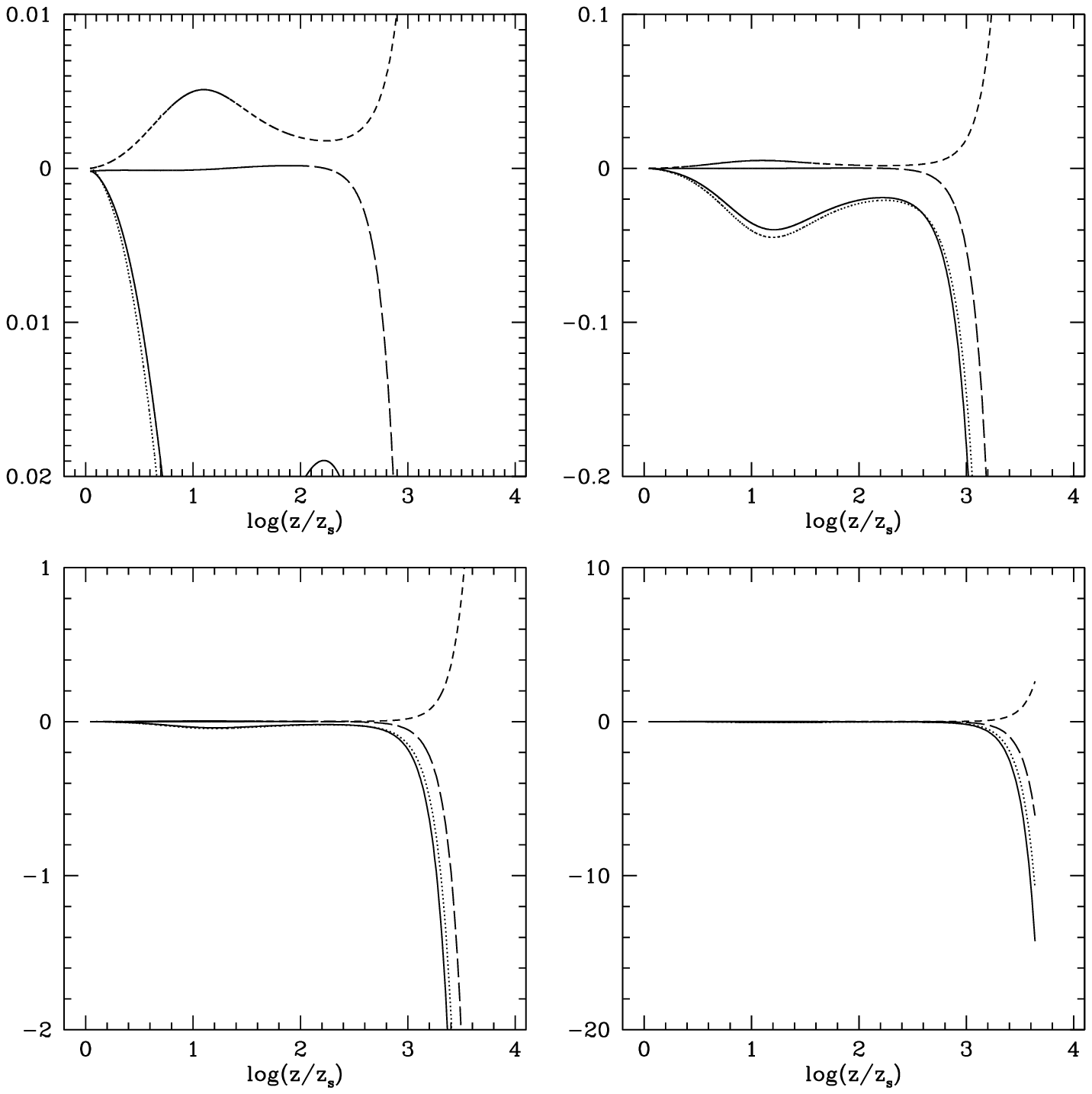}
\caption{
Local solution existence for the 1D vertical wind model with the flux
distribution of a standard Shakura-Sunyaev disk at
$r=20 \, r_i$:
a local solution exists in the vicinity of a critical point
$z$
if the expression
$\beta''\, \dot{m} \, (\omega')^2 + \beta' \, \dot{m} \, \omega'' - n''$
(solid curve)
is negative.
This figure shows that all points ranging from
$\approx 1.107$
times the sonic height to beyond 1000 times the sonic height fulfill
this condition.
From
$\approx 1.107$
times the sonic height down to the sonic point,
the critical point conditions cannot be met
(see Figure~\ref{fig_pkb21_20_n}).
In this plot the terms
$\beta'' \, \dot{m} \, (\omega')^2$
(dotted curve),
$\beta' \, \dot{m} \, \omega''$
(short dash curve),
and
$-n''$
(long dash curve)
are also shown.
The physical parameters used are the same as in
Figure~\ref{fig_pkb21_02_h}.
}
\label{fig_pkb21_20_l}
\end{figure}

\begin{figure}
\epsscale{1.0}
\plotone{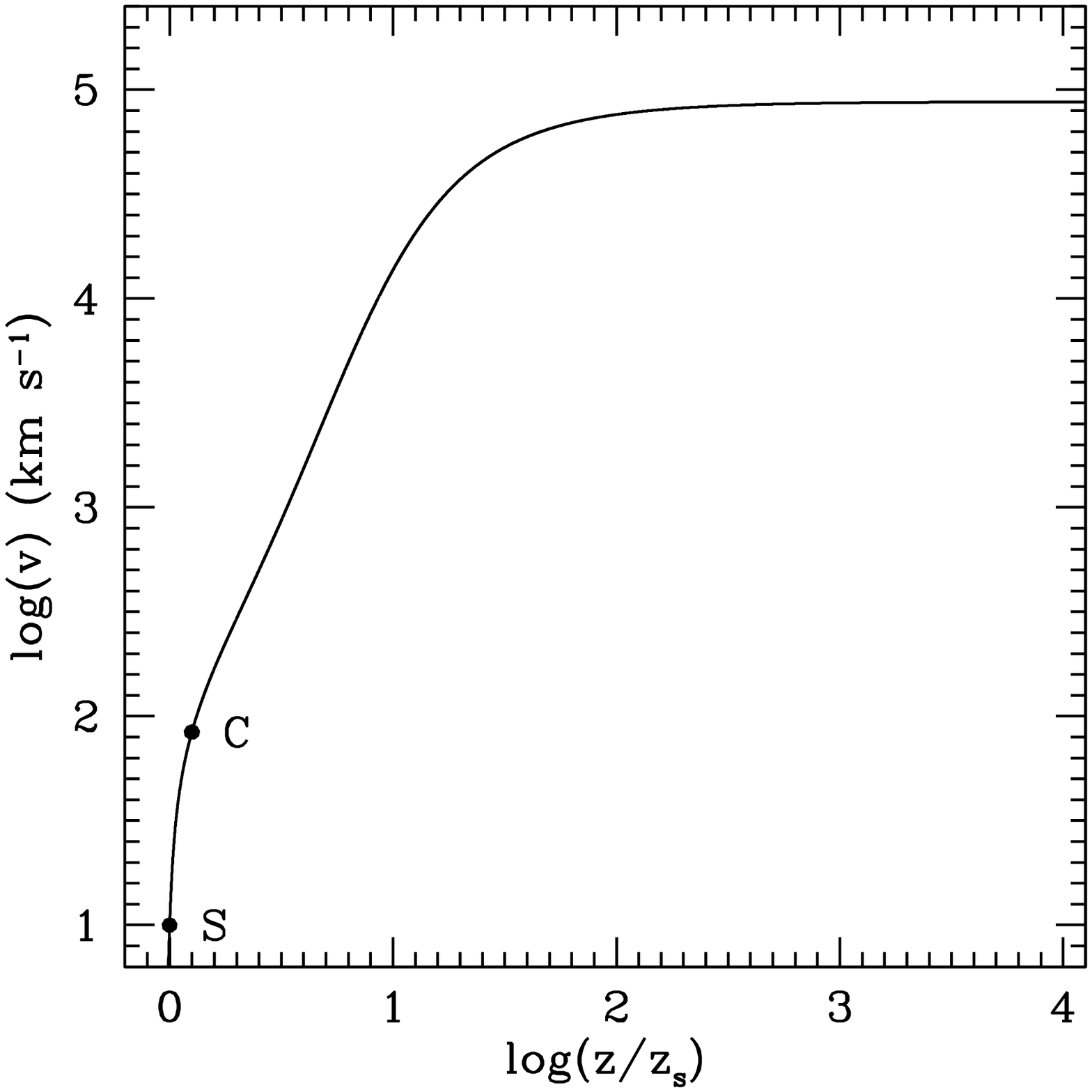}
\caption{
Velocity vs. position for the 1D vertical wind model with the flux
distribution of a standard Shakura-Sunyaev disk at
$r=20 \, r_i$.
The critical point position is determined with the condition that,
upon integration of the equation of motion,
the correct sonic point position is found.
``C''
indicates the critical point and
``S''
indicates the sonic point.
The physical parameters used are the same as in
Figure~\ref{fig_pkb21_02_h}.
}
\label{fig_pkb21_20_v}
\end{figure}

\begin{figure}
\epsscale{0.95}
\plotone{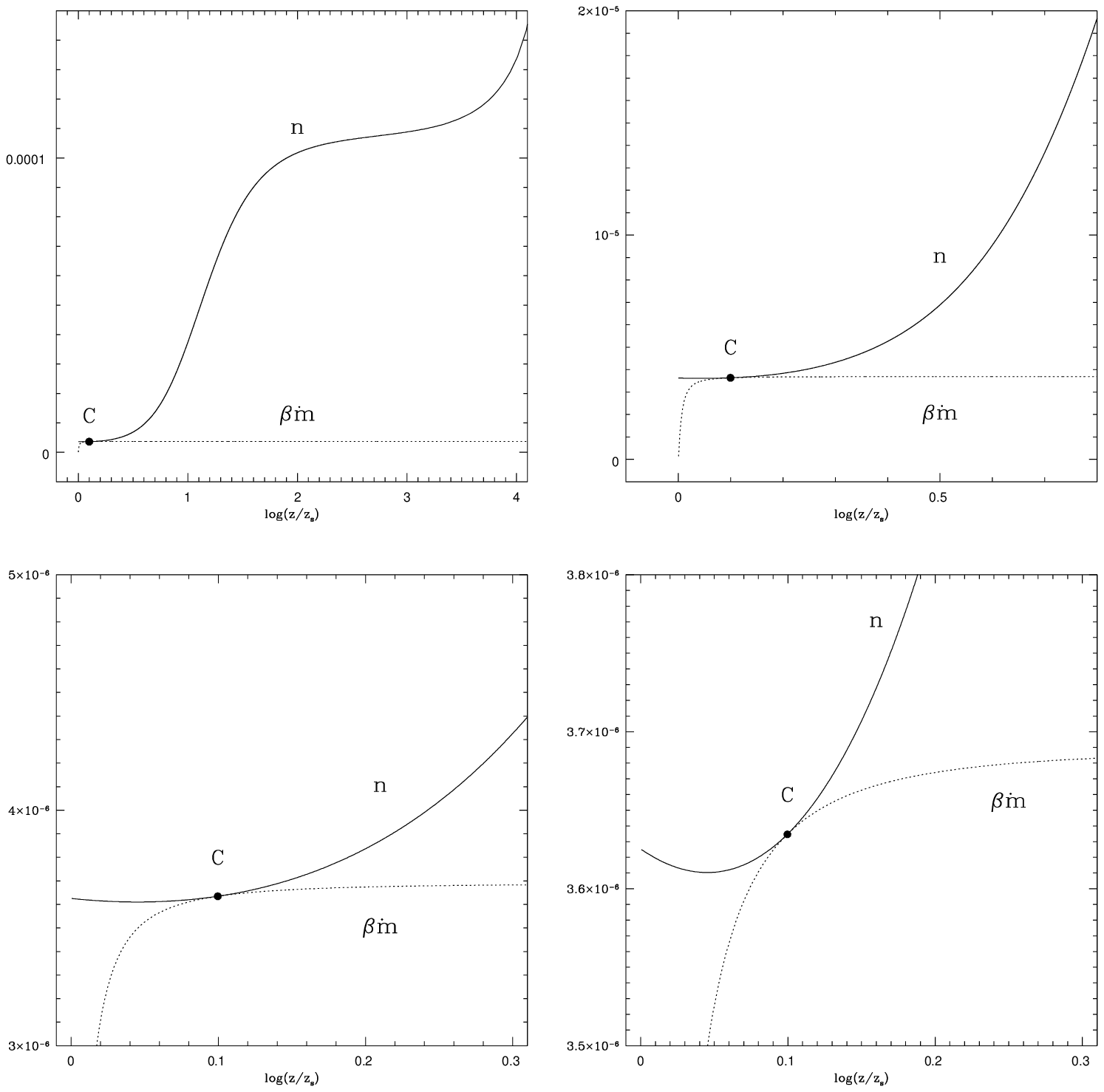}
\caption{
Necessary condition for the global solution existence for the 1D
vertical wind model with the flux distribution of a standard
Shakura-Sunyaev disk at
$r=20 \, r_i$
for the critical point shown in
Figure~\ref{fig_pkb21_20_v}:
upon the integration of the equation of motion,
it must hold that
$\beta(\omega) \, \dot{m} < n(q)$
at points other than the critical point
[when
 $h(q)<0$
 and the wind is supersonic],
and
$\beta(\omega) \, \dot{m} = n(q)$
at the critical point.
Presented here is the nozzle function
$n$
(solid curve)
and the
$\beta \, \dot{m}$
function
(dotted curve)
vs. position.
``C''
indicates the critical point.
The physical parameters used are the same as in
Figure~\ref{fig_pkb21_02_h}.
}
\label{fig_pkb21_20_nb}
\end{figure}

\begin{figure}
\epsscale{1.0}
\plotone{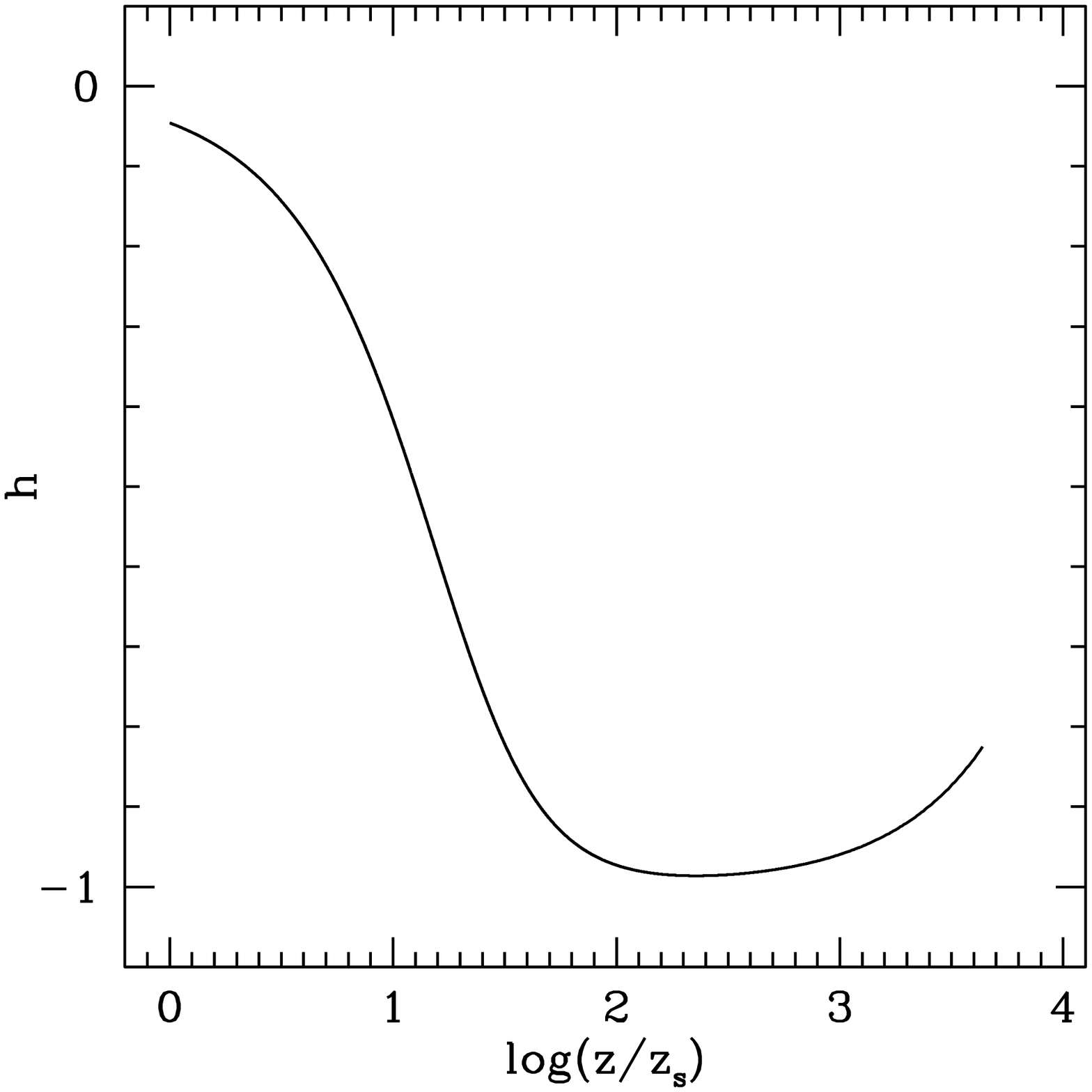}
\caption{
The
$h$
function for the 1D vertical wind model with the flux distribution of a
standard Shakura-Sunyaev disk at
$r=50 \, r_i$.
A necessary requirement for the critical point is that the
$h$
function must be negative at that point.
This figure shows that all points ranging from the sonic point to beyond
1000 times the sonic height fulfill this condition.
The physical parameters used are the same as in
Figure~\ref{fig_pkb21_02_h}.
}
\label{fig_pkb21_50_h}
\end{figure}

\begin{figure}
\epsscale{1.0}
\plotone{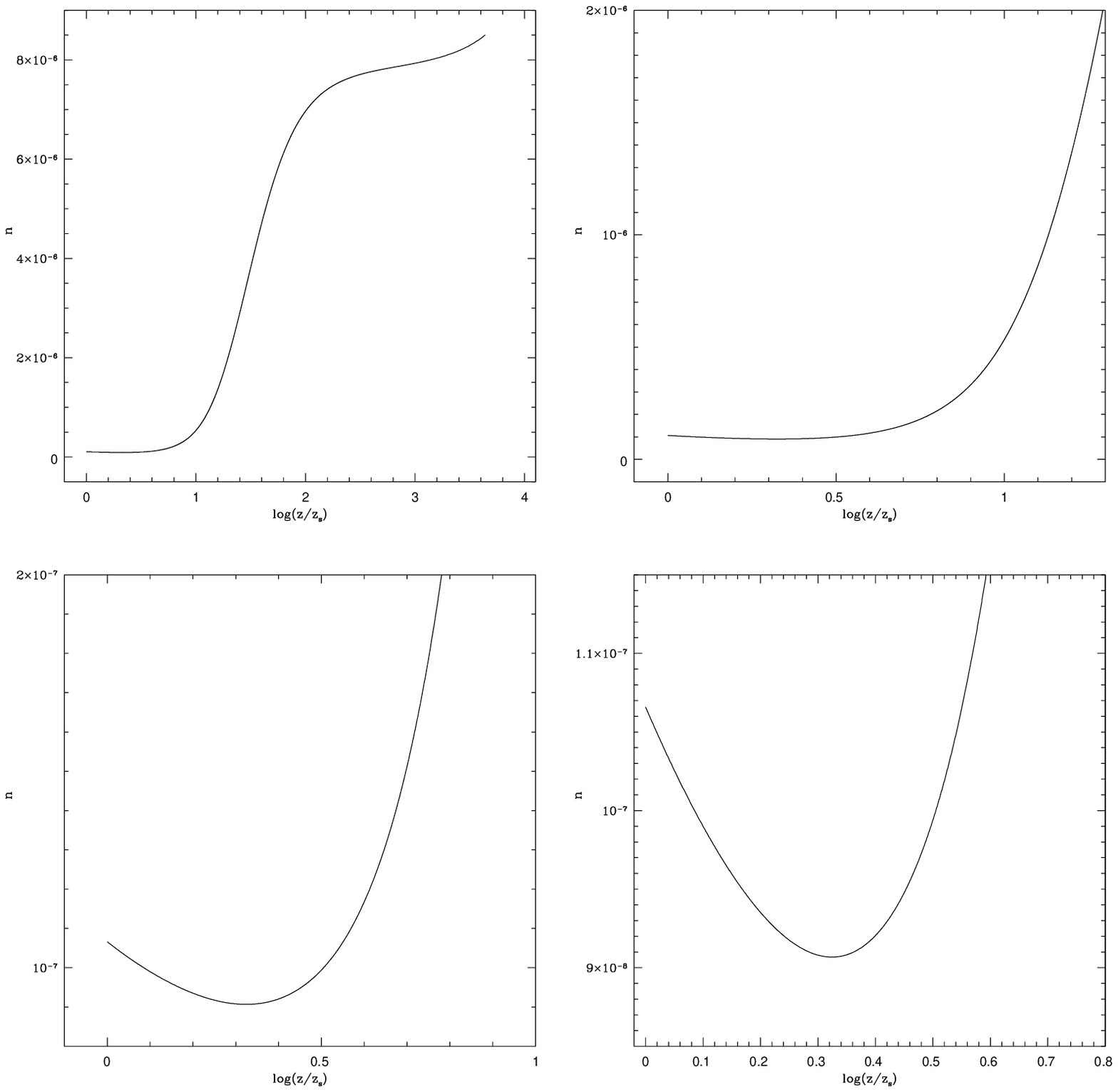}
\caption{
The nozzle function
$n$
for the 1D vertical wind model with the flux distribution of a standard
Shakura-Sunyaev disk at
$r=50 \, r_i$.
A necessary requirement for the critical point in an isothermal
line-driven wind is that the nozzle function must be locally increasing 
(i.e.,
 $dn/dq > 0$)
at that point.
This figure shows that all points ranging from
$\approx 2.11$
times the sonic height to beyond 1000 times the sonic height fulfill
this condition.
However,
it is also shown that points ranging from the sonic point up to
$\approx 2.11$
times the sonic height do not fulfill the condition.
The physical parameters used are the same as in
Figure~\ref{fig_pkb21_02_h}.
}
\label{fig_pkb21_50_n}
\end{figure}

\begin{figure}
\epsscale{0.95}
\plotone{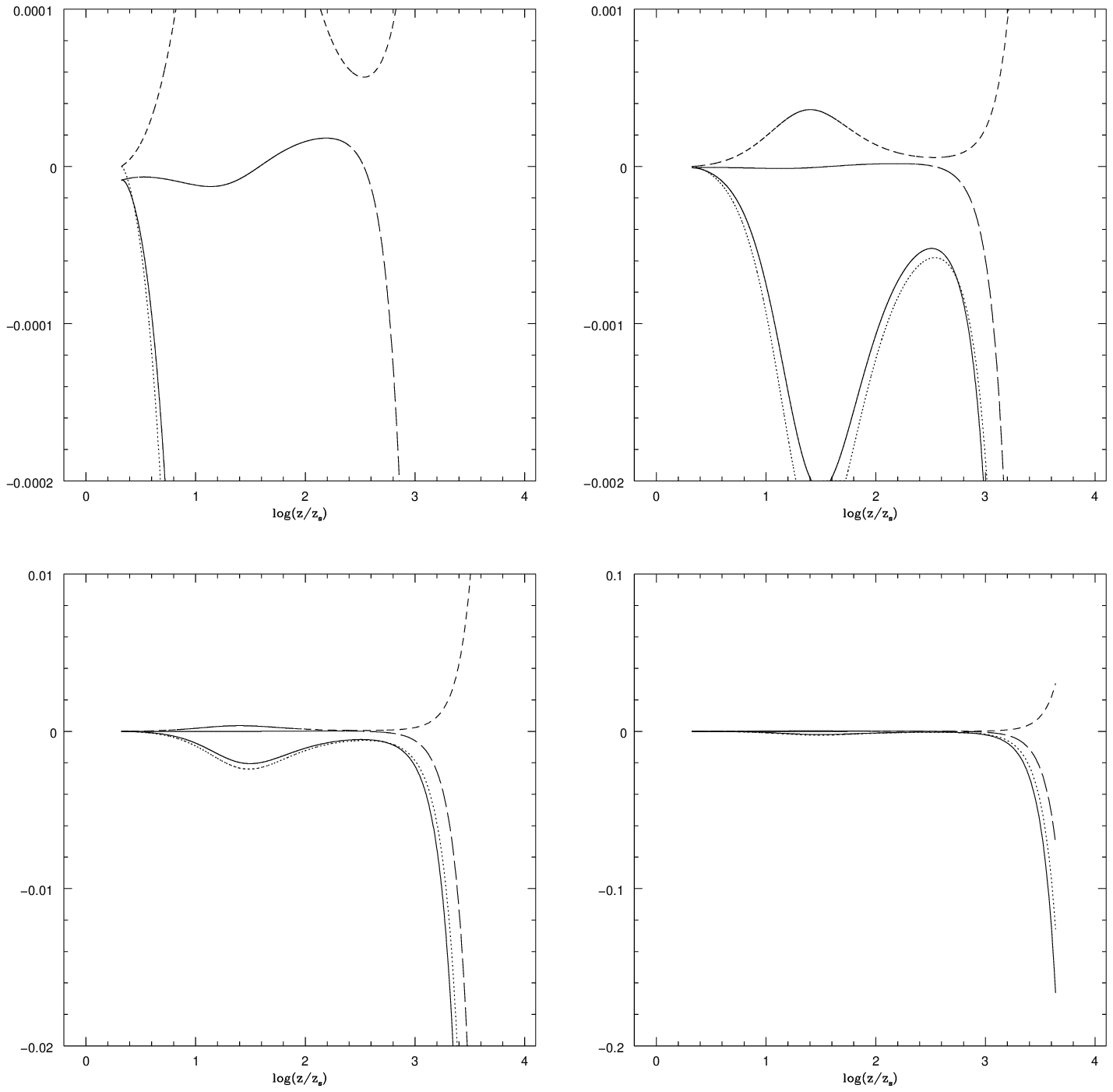}
\caption{
Local solution existence for the 1D vertical wind model with the flux
distribution of a standard Shakura-Sunyaev disk at
$r=50 \, r_i$:
a local solution exists in the vicinity of a critical point
$z$
if the expression
$\beta''\, \dot{m} \, (\omega')^2 + \beta' \, \dot{m} \, \omega'' - n''$
(solid curve)
is negative.
This figure shows that all points ranging from
$\approx 2.11$
times the sonic height to beyond 1000 times the sonic height fulfill
this condition.
From
$\approx 2.11$
times the sonic height down to the sonic point,
the critical point conditions cannot be met
(see Figure~\ref{fig_pkb21_50_n}).
In this plot the terms
$\beta'' \, \dot{m} \, (\omega')^2$
(dotted curve),
$\beta' \, \dot{m} \, \omega''$
(short dash curve),
and
$-n''$
(long dash curve)
are also shown.
The physical parameters used are the same as in
Figure~\ref{fig_pkb21_02_h}.
}
\label{fig_pkb21_50_l}
\end{figure}

\begin{figure}
\epsscale{1.0}
\plotone{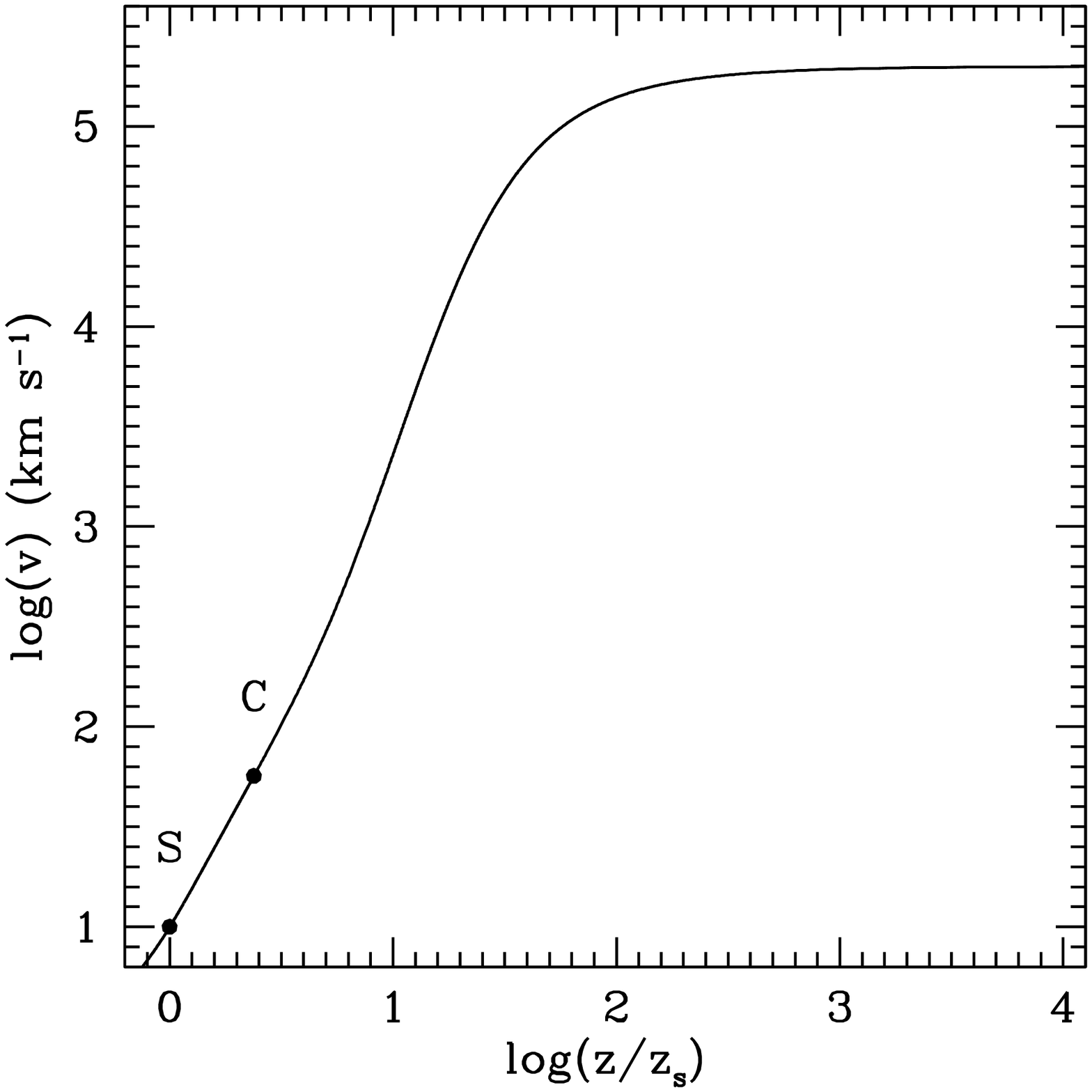}
\caption{
Velocity vs. position for the 1D vertical wind model with the flux
distribution of a standard Shakura-Sunyaev disk at
$r=50 \, r_i$.
The critical point position is determined with the condition that,
upon integration of the equation of motion,
the correct sonic point position is found.
``C''
indicates the critical point and
``S''
indicates the sonic point.
The physical parameters used are the same as in
Figure~\ref{fig_pkb21_02_h}.
}
\label{fig_pkb21_50_v}
\end{figure}

\begin{figure}
\epsscale{1.0}
\plotone{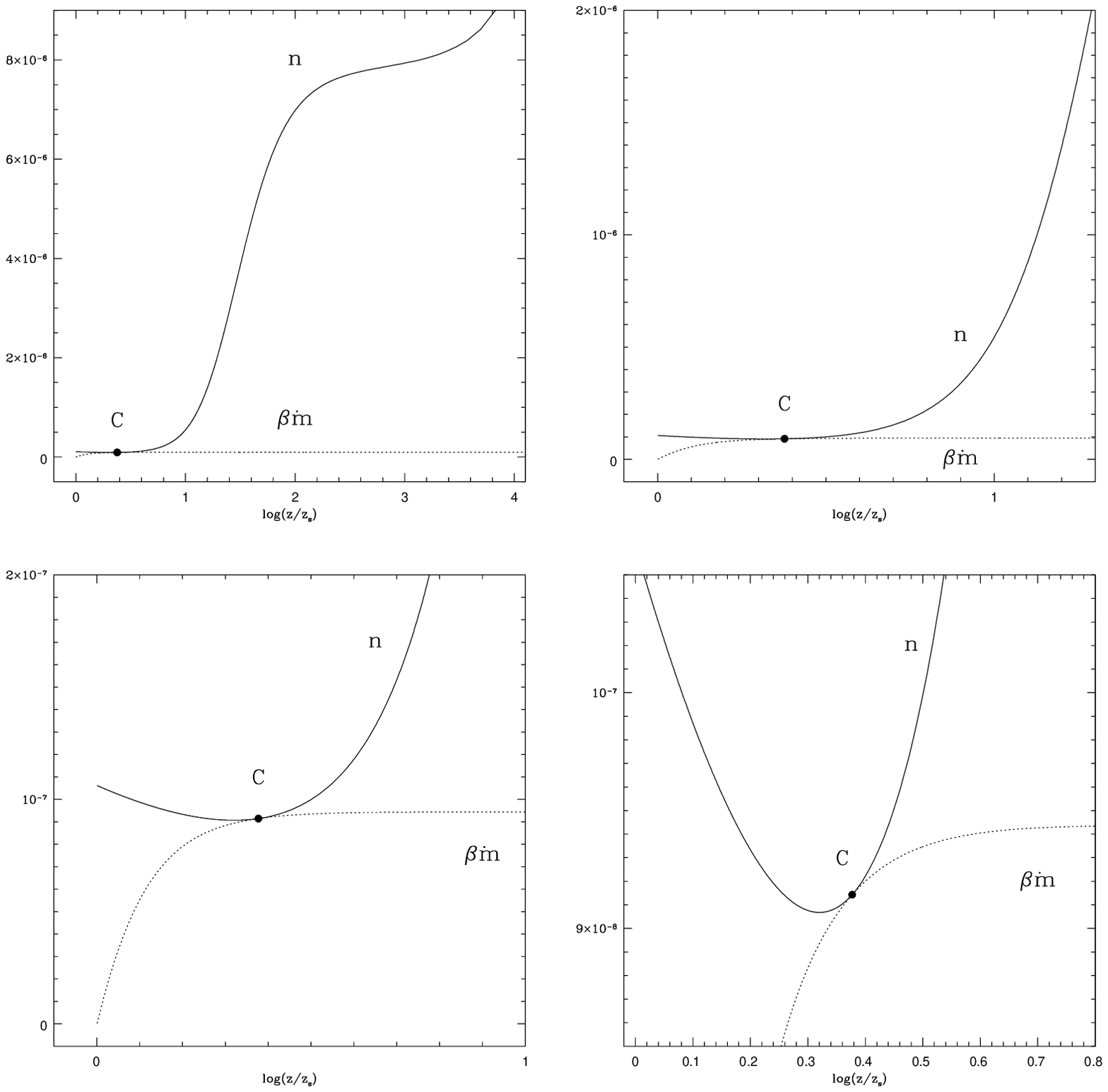}
\caption{
Necessary condition for the global solution existence for the
1D vertical wind model with the flux distribution of a standard
Shakura-Sunyaev disk at
$r=50 \, r_i$
for the critical point shown in Figure~\ref{fig_pkb21_50_v}:
upon the integration of the equation of motion,
it must hold that
$\beta(\omega) \, \dot{m} < n(q)$
at points other than the critical point
[when
 $h(q)<0$
 and the wind is supersonic],
and
$\beta(\omega) \, \dot{m} = n(q)$
at the critical point.
Presented here is the nozzle function
$n$
(solid curve)
and the
$\beta \, \dot{m}$
function
(dotted curve)
vs. position.
``C''
indicates the critical point.
The physical parameters used are the same as in
Figure~\ref{fig_pkb21_02_h}.
}
\label{fig_pkb21_50_nb}
\end{figure}

Figures~\ref{fig_pkb21_10_h}-\ref{fig_pkb21_10_nb},
Figures~\ref{fig_pkb21_20_h}-\ref{fig_pkb21_20_nb},
and Figures~\ref{fig_pkb21_50_h}-\ref{fig_pkb21_50_nb} plot the results
from an equivalent analysis to wind streamlines starting at
$r = 10 \, r_i$,
$r = 20 \, r_i$,
and $r = 50 \, r_i$ respectively.
The results show that steady line-driven disk wind solutions exist for
a standard Shakura-Sunyaev disk.

A significant result we find is that the critical point tends to be
closer to the disk surface in the inner disk regions,
and farther out from the disk surface in the outer disk regions.
Physically this is due to the length scales in these different regions.
The disk wind,
at each radii,
results from the balancing of the gravitational forces and the
radiation pressure forces.
This balance can be presented quantitatively by the nozzle function
defined in Paper~I and Paper~II.
The spatial dependence of the nozzle function in turn is crucial in
determining the critical point position.
As larger disk radii are considered,
the scale length of the corresponding nozzle function also increases,
and therefore the critical point position increases as well.

Also,
the resulting local wind mass loss rates,
for the different radii considered,
are in good agreement with the 2.5D numerically intensive CV disk wind
models of
\cite{per97a,per00};
for example for a radius of $10r_i$ the local wind mass loss rates
found in this work is
$5.0 \times 10^{-7} {\rm g} \, {\rm s}^{-1} \, {\rm cm}^{-2}$,
and for the same physical model parameters the previous numerically
intensive 2.5D calculations estimated a value of
$7.1 \times 10^{-7} {\rm g} \, {\rm s}^{-1} \, {\rm cm}^{-2}$
for the same radius.
By confirming the overall earlier results of Pereyra and collaborators,
we conclude that the likely scenario for the wind outflows in CVs is
a line-driven disk wind.
This scenario was suggested for CVs early on by \citet{cor82}.
\citet{per97a,per00} showed that the line-driven disk wind scenario
was able to account for the general forms of the \ion{C}{4}
$\lambda\lambda$1549 line profile and its general dependence on
viewing angle
[for a review on CV properties in general, see \citet{war95}].

The steady nature of line-driven disk wind that we find here may also
be relevant to QSO studies. \citet{tur84} had suggested this scenario
early-on based on observational constraints.
The first serious attempt to study this scenario was by
\citet{mur95},
who developed 1D streamline models and found that this scenario could
account for several QSO observational features,
such as the terminal wind velocities inferred from
broad absorption lines (BALs)
(when present)
and the general form of single-trough BALs
(when present).
As discussed in the Introduction,
an additional constraint on the BAL region,
is that it presents a steady velocity structure.
Our results here show that the line-driven disk wind scenario is
consistent with the observed steady velocity structure.

Thus both observational constraints and theoretical/computational
studies to date seem to indicate that the line-driven disk wind
scenario is a promising one to account for the BALs commonly
observed in QSOs.

In future work we plan to study several aspects of this scenario and
develop more realistic models.
 
\section{Summary and Conclusions}
\label{sec_sumcon}

In Paper~I we had shown that steady wind solutions can exist by using
``simple'' models that mimic the disk environment.
These models are more readily analyzable than the more detailed models
presented here.
In Paper~II we extended the concepts of Paper~I,
and discussed in detail aspects of the steady/unsteady wind analysis
that was presented in Paper~I.
The objective of this work is to determine,
in a manner independent of the results of previous
numerically-intensive 2.5D hydrodynamic simulations,
whether steady line-driven disk wind solutions exist under the flux
distribution of a standard disk \citep{sha73}.

Our main conclusion pertaining to the more realistic models presented
here is that a line-driven wind,
arising from a steady disk using the flux distribution of a standard
Shakura-Sunyaev disk model,
is steady.
As we had discussed in Paper~II,
when including gas pressures effects,
the spatial dependence of the nozzle function continues to play a key
role in determining the steady/unsteady nature of supersonic line-driven
wind solutions.

These results are consistent with the steady nature of the 1D
streamline disk wind models of Murray and collaborators
\citep{mur95,mur96,chi96,mur98}.
These results also confirm the results of the 2.5D steady disk wind
models of Pereyra and collaborators
\citep{per97a,per00,hil02,per03}.

Another result that we find is that the critical point of the wind tends
to be closer to the disk surface as smaller radii closer to the inner
disk region are considered.

As we develop more realistic models,
we aim towards placing stronger constraints on the accretion disk
wind scenario,
particularly as they apply to QSOs.
If the scenario proves tenable,
then we may have a well-defined route towards a better understanding
of these astronomically fundamental objects.

\acknowledgments

We wish to thank Kenneth G. Gayley and Norman W. Murray
for many useful discussions.
This work is supported by the National Science
Foundation under Grant AST-0071193,
and by the National Aeronautics and Space Administration
under Grant ATP03-0104-0144.


\clearpage

\clearpage

\end{document}